\documentclass[
preprint,
superscriptaddress,
preprintnumbers,
nofootinbib,
notitlepage,
amsmath,amssymb,
aps,
pre,
longbibliography,
floatfix
]{revtex4-1}

\usepackage[english]{babel} 
\usepackage[latin9]{inputenc}
\usepackage{xcolor}
\usepackage{float}
\usepackage{graphicx}
\usepackage{units}
\usepackage{bm}
\usepackage{array,overpic}

\usepackage{hyperref}
\hypersetup{
    breaklinks=true
}
\usepackage{rotating}
 \usepackage{soul}
\soulregister\ref7

\allowdisplaybreaks

\newcommand{\ri}{\mathrm{i}}
\newcommand{\rd}{\mathrm{d}}

\newcommand{\De}{\textrm{De}}   
\newcommand{\Pen}{\textrm{Pe}}  

\definecolor{bluegray}{rgb}{0.4, 0.6, 0.8}
\definecolor{cadmiumorange}{rgb}{0.91, 0.41, 0.17}

\usepackage{todonotes}

\begin{document}
\setlength{\parskip}{0pt}
\title{Time-averaged transport in oscillatory squeeze flow of a viscoelastic fluid}

\author{Rui Yang}
 \affiliation{Department of Civil and Environmental Engineering, Technion -- Israel Institute of Technology, Haifa 32000, Israel}
\author{Ivan C. Christov}
 \affiliation{School of Mechanical Engineering, Purdue University,
West Lafayette, Indiana 47907, USA}
\author{Ian M. Griffiths}
 \affiliation{Mathematical Institute, University of Oxford, Radcliffe Observatory Quarter, Oxford OX2 6GG, United Kingdom}
\author{Guy Z. Ramon}\thanks{Author to whom correspondence should be addressed.}
 \email{ramong@technion.ac.il}
 \affiliation{Department of Civil and Environmental Engineering, Technion -- Israel Institute of Technology, Haifa 32000, Israel}


\date{\today}

\begin{abstract}
Periodically-driven flows are known to generate non-zero, time-averaged fluxes of heat or solute species, due to the interactions of out-of-phase velocity and temperature/concentration fields, respectively. Herein, we investigate such transport (a form of the well-known Taylor--Aris dispersion) in the gap between two parallel plates, one of which oscillates vertically, generating a time-periodic squeeze flow of either a newtonian or Maxwellian fluid. Using the method of multiple time-scale homogenization, the mass/heat balance equation describing transport in this flow is reduced to a one-dimensional advection--diffusion--reaction equation. This result indicates three effective mechanisms in the mass/heat transfer in the system: an effective diffusion that spreads mass/heat along the concentration/temperature gradient, an effective advective flux, and an effective reaction that releases or absorbs mass/heat - in the time-averaged frame. Our results demonstrate that there exist resonant modes under which the velocity peaks when the dimensionless plate oscillation frequency (embodied by the Womersley number, the ratio of the transient inertia to viscous forces) approaches specific values. As a result, transport in this flow is significantly influenced by the dimensionless frequency. On the one hand, the effective, time-averaged dispersion coefficient is always larger than the molecular diffusivity, and is sharply enhanced near resonance. The interaction between fluid elasticity and the oscillatory forcing enhances the efficiency of transport in the system. On the other hand, the identified effective advection and reaction mechanisms may transport mass/heat from regions of high concentration/temperature to those of low concentration/temperature, or vice versa, depending on the value of dimensionless frequency. Ultimately, it is shown that the oscillatory squeeze flow can either enhance or diminish transport, depending on the interplay of these three effective (homogenized) mechanisms.  
\end{abstract}

\maketitle

\section{Introduction}

The spreading of a scalar in a flow due to the combined action of diffusion, advection and reaction is widely known as Taylor--Aris (shear) dispersion~\citep{taylor_dispersion_1953,aris_dispersion_1960,sankarasubramanian1973unsteady,PhysRevFluids.4.034501,Young19991}. Shear dispersion has been studied extensively, in particular, for flows that are oscillatory in time \cite{chatwin_longitudinal_1975-1,watson_diffusion_1983,joshi_experimental_1983,vedel_hovad_bruus_2014,gill1971dispersion,smith1982contaminant}, due to their relevance to transport in arteries~\citep{gentile_transport_2008}, pulmonary airways~\citep{fredberg_augmented_1980,eckmann_grotberg_1988,Grotberg1994}, the cerebrospinal fluid \cite{PhysRevFluids.5.043102}, through bones \cite{SCHMIDT20052337}, liquid membranes \cite{Leighton1988} and wave boundary layers \citep{mei_dispersion_1994}, actuation of oscillatory flow via electro-osmotic forces \citep{ramon_solute_2011,Ramon2018}, and so on. The analogous heat-transfer problem -- generating a considerable augmentation of the heat flux through flow oscillations -- has also been observed (see, for example, \cite{kurzweg_heat_1984,kurzweg_enhanced_1985-1,lambert_heat_2009}). 

\begin{figure}
    \centering
    \begin{overpic}[scale=0.8,tics=5]{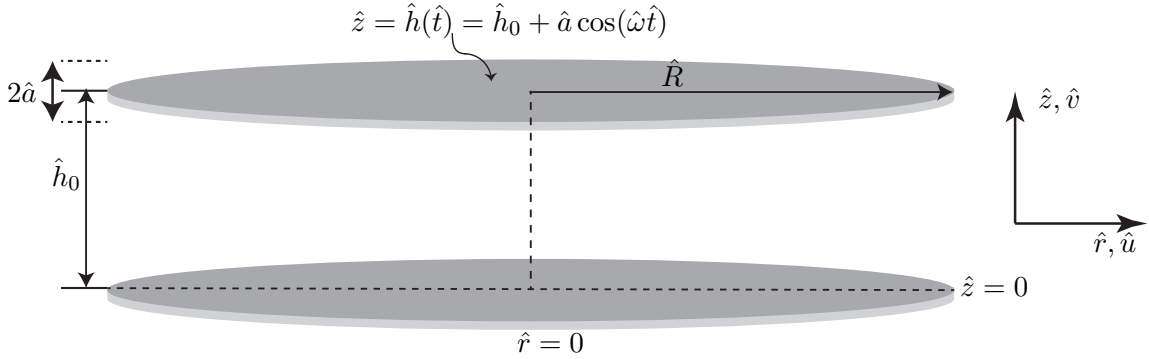}
\put(-3,20.6){$2 \hat{a}$}
\put(56,22){$\hat{R}$}
\put(1,13){$\hat{h}_0$}
\put(28,27){$\hat{z}=\hat{h}(\hat{t})=\hat{h}_{0}+\hat{a} \cos (\hat{\omega} \hat{t})$}
\put(43,-2){$\hat{r}=0$}
\put(83,3){$\hat{z}=0$}
\put(90,20){$\hat{z},\hat{v}$}
\put(95,7){$\hat{r},\hat{u}$}
\end{overpic}
    \caption{Schematic of the oscillatory squeeze flow in cylindrical coordinates.}
    \label{fig:schematic}
\end{figure}

In what follows, we consider an oscillatory squeeze flow (OSF), in which a viscoelastic fluid is driven periodically by the motion of one of the confining, parallel planes (see Fig.~\ref{fig:schematic} for a schematic of the studied configuration). This type of setting exists in some devices using magnetorheological fluids \citep{li_benchmark_2018} or electrorheological fluids \citep{wingstrand_oscillatory_2016-1}, where the fluid is squeezed periodically under variable magnetic or electric field, as means of achieving continuously variable control of mechanical vibrations, and also envisioned as a mode of achieving control over transport in confined systems.

An early theoretical analysis of OSF was conducted by Phan-Thien and Tanner \citep{phan-thien_small_1980,phan-thien_viscoelastic_1983}, who provided analytical solutions for the velocity profile and the normal force required to drive the plate. They proposed that the dynamic properties of polymeric liquids could be measured using a plastometer operating in the OSF mode. Since then, OSF equipment has been proposed and widely used in the rheology community for evaluating the processability of polymer melts, and for possible macromolecular characterization \citep{field_experimental_1996,engmann_squeeze_2005}. Furthermore, its ability to augment transport of mass and/or heat serves as a motivation for understanding the characteristics of this system.

Studies by \citet{phan-thien_small_1980} and \citet{bell_oscillatory_2006} demonstrated that a periodic flow in the streamwise direction (perpendicular to the oscillation direction of the plates) of an OSF exhibits a time dependence similar to an oscillatory pipe flow. Hence, there is reason to believe that shear-induced dispersion should also exist in the OSF, and dispersion would augment the mixing of the solute and the bulk flow along the streamwise direction. In this vein, \citet{stone_dispersion_1999} investigated the shear dispersion of a solute in a steady radial flow between two parallel plates, and discovered a radially-dependent effective diffusivity in the streamwise direction. \citet{creech_dispersive_2001} proposed a model to analyze dispersion in a specific case of the OSF, which occurs in the tear layer of the eye, sandwiched between the cornea and a soft contact lens. They found that the dispersion coefficient is orders of magnitude higher than the molecular diffusivity, facilitating mixing between the cornea and a soft contact lens. 
Specifically, the dispersion-augmented mixing is much faster than that which can be achieved by diffusion alone. 
{However, the model neglects the advective contribution of streamwise velocity variations, while using a form of the velocity field valid for low frequencies. These simplifications limit the applicability of the study.} In the present study, we relax these assumptions, while also extending it so as to the visco-elastic character of the fluid and the effect on Taylor--Aris dispersion.

In this paper, in view of the mathematical and physical analogy between mass diffusion and heat conduction, we investigate the dispersion in both contexts via a unified approach. We consider a domain confined by two parallel plates that enclose a Newtonian or a viscoelastic Maxwell fluid. One of the plates oscillates vertically in a sinusoidal fashion, squeezing the enclosed fluid. We determine the fluid velocity field and reduce the mass/heat balance equation into a one-dimensional advection--diffusion--reaction equation via the asymptotic method of multiple-scales homogenization. Interestingly, our results show that the OSF can either enhance or diminish mass/heat transfer, depending on the interplay of the effective diffusion, advection and reaction mechanisms. To this end, this paper is organized as follows: We begin by formulating the problem of the viscoelastic OSF between two plates. First, we obtain an analytical solution for the velocity profile in \textsection\ref{sec:Velocity-field}. Then, in \textsection\ref{sec:Diffusion}, we combine the latter with the effective advection--diffusion-reaction equation for transport in the OSF. We present the results and discussion related to the velocity field in \textsection\ref{sec:dis_velocity} and the transport characteristics in \textsection\ref{sec:dis_diffusion}. Concluding remarks are given in \textsection\ref{sec:Conclusion}. Appendices are included with a detailed derivation of the homogenized transport equation, as well as a calculation of the power required to drive the system.

\section{Model formulation\label{sec:Problem-description}}

We consider the flow and scalar transport in a viscoelastic fluid flow confined between two concentric discs with the same radii of $\hat{R}$, separated by a distance $\hat{h}_{0}$ (see schematic in Fig.~\ref{fig:schematic}).  The upper disc oscillates sinusoidally with amplitude $\hat{a}$ about a mean position $\hat{h}_0$, such that its position in time is $\hat{z} = \hat{h}(\hat{t})$, and the distance between the two discs is
\begin{equation}
\hat{h}(\hat{t})=\hat{h}_{0}+\hat{a}\cos(\hat{\omega} \hat{t}),
\end{equation}
where $\hat{\omega}$ is the angular frequency, and $\hat{t}$ is time. The vertical velocity of the top disc is therefore 
\begin{equation}
\hat{h}'(\hat{t})=-\hat{a}\hat{\omega}\sin(\hat{\omega} \hat{t})=\Re\left\{ \hat{a}\hat{\omega}\mathrm{i}\mathrm{e}^{\mathrm{i}\hat{\omega} \hat{t}}\right\},
\label{eq:velocity of plate}
\end{equation}
where $\Re\{\,\cdot\,\}$ denotes the real part of a complex quantity, and primes denote differentiation.

\subsection{Velocity field}
\label{sec:Velocity-field}

\subsubsection{Governing equations}

We begin with the equations of motion for an incompressible flow. The continuity equation is 
\begin{equation}
\frac{1}{\hat{r}}\frac{\partial}{\partial \hat{r}}\left(\hat{r}\hat{u}\right)+\frac{\partial \hat{v}}{\partial \hat{z}}=0,
\label{eq:continuity}
\end{equation}
where the two-dimensional velocity field is $\hat{\bm{u}}=\hat{u}\bm{e}_r+\hat{v}{\bm{e}_z}$, with $\bm{e}_r$ and ${\bm{e}_z}$ the unit vectors in the $\hat{r}$ and $\hat{z}$ directions, respectively (see Fig.~\ref{fig:schematic}),
and the momentum equation, neglecting body forces, is
\begin{equation}
\hat{\rho}\frac{\mathrm{D}\hat{\bm{u}}}{\mathrm{D} \hat{t}}=-\nabla \hat{p}+\nabla\cdot\hat{\boldsymbol{\tau}},
\label{eq:momentum_0}
\end{equation}
where $\hat{\boldsymbol{\tau}}$ is the (deviatoric) viscous stress tensor, $\hat{\rho}$ is the density of the fluid, and $\hat{p}$ is the pressure.
The constitutive equation for the incompressible upper-convected Maxwell fluid (see, for example, \cite{phan-thien_viscoelastic_1983}) is
\begin{equation}
\hat{\boldsymbol{\tau}} + \hat{\lambda}_0\left(\frac{\mathrm{D}\hat{\boldsymbol{\tau}}}{\mathrm{D}\hat{t}}-\left[\hat{\boldsymbol{\tau}}\cdot\left(\nabla\hat{\bm{u}}\right)+\left(\nabla\hat{\bm{u}}\right)^{T}\cdot\hat{\boldsymbol{\tau}}\right]\right)=\hat{\mu}\left(\nabla\hat{\bm{u}}+\nabla\hat{\bm{u}}^{T}\right),
\label{eq:maxwell}
\end{equation}
where $\hat{\lambda}_0$ is the viscoelastic relaxation time, and $\hat{\mu}$ is the usual, Newtonian,  dynamic shear viscosity.

In what follows we shall assume that the amplitude of the disc oscillations is small, i.e., the dimensionless displacement amplitude of the upper plate $\delta={\hat{a}}/{\hat{h}_{0}}\ll1$, while the frequency is unrestricted. With this assumption, combining Eq.~(\ref{eq:maxwell}) and Eq.~(\ref{eq:momentum_0}) results in (see Appendix \ref{app:DDt} for derivation)
\begin{subequations}\begin{align}
\hat{\lambda}_0\frac{\partial^{2}\hat{u}}{\partial \hat{t}^{2}} + \frac{\hat{\lambda}_0}{\hat{\rho}}\frac{\partial^{2}\hat{p}}{\partial \hat{t}\partial \hat{r}}+\frac{\partial \hat{u}}{\partial \hat{t}}&=-\frac{1}{\hat{\rho}}\frac{\partial \hat{p}}{\partial \hat{r}}+\frac{\hat{\mu}}{\hat{\rho}}\left(\frac{\partial^{2}\hat{u}}{\partial \hat{r}^{2}}+\frac{1}{\hat{r}}\frac{\partial \hat{u}}{\partial \hat{r}}+\frac{\partial^{2}\hat{u}}{\partial \hat{z}^{2}}-\frac{\hat{u}}{\hat{r}^{2}}\right),
\label{eq:momentum1_dim}
\\
\hat{\lambda}_0\frac{\partial^{2}\hat{v}}{\partial \hat{t}^{2}} + \frac{\hat{\lambda}_0}{\hat{\rho}}\frac{\partial^{2}\hat{p}}{\partial \hat{t}\partial \hat{z}}+\frac{\partial \hat{v}}{\partial \hat{t}}&=-\frac{1}{\hat{\rho}}\frac{\partial \hat{p}}{\partial \hat{z}}+\frac{\hat{\mu}}{\hat{\rho}}\left(\frac{\partial^{2}\hat{v}}{\partial \hat{r}^{2}}+\frac{1}{\hat{r}}\frac{\partial \hat{v}}{\partial \hat{r}}+\frac{\partial^{2}\hat{v}}{\partial \hat{z}^{2}}\right).
\label{eq:momentum2_dim}
\end{align}\end{subequations}

We impose no slip and no penetration at the solid walls, which correspond to
\begin{equation}
\hat{u}(0)=0,\qquad \hat{u}(\hat{h})=0,\qquad \hat{v}(\hat{h})=\Re\left\{ \hat{a}\hat{\omega}\mathrm{i}\mathrm{e}^{\mathrm{i}\hat{\omega} \hat{t}}\right\} ,\qquad \hat{v}(0)=0.\label{eq:original BC_1palte0}
\end{equation}
In principle, we must also specify conditions at $\hat{r}=\hat{R}$ to close the problem. We discuss this in more detail in \textsection\ref{Section:Solution velocity field} below.

\subsubsection{Non-dimensionalization}
We posit lubrication-type scalings with the pressure made dimensionless using the viscous stress scale: 
\begin{equation}
\hat{u}= \hat{U}{u},\quad \hat{v}=\epsilon \hat{U}{v},\quad \hat{t}= \hat{T}{t},\quad \hat{r}= \hat{R}{r},\quad \hat{z}=\epsilon \hat{R}{z},\quad \hat{p}=\frac{\hat{\mu} \hat{U}}{\epsilon^{2}\hat{R}}{p},\quad \hat{h}=\hat{h}_0 h.
\label{eq:nond_notation}
\end{equation}
Here, $\epsilon= \hat{h}_0/\hat{R}$. 
The vertical velocity scale $\epsilon \hat{U}$ is set by the oscillations of the wall, hence we have $\hat{U}=\hat{a}\hat{\omega}/\epsilon$.
The natural time scale of the problem is the inverse of the oscillation frequency of the wall, i.e., $\hat{T}$ = 1/$\hat{\omega}$. The dimensionless versions of the momentum equations~\eqref{eq:momentum1_dim} and \eqref{eq:momentum2_dim} are then
\begin{subequations}\begin{align}
\De\,\alpha^{2}\frac{\partial^{2}{u}}{\partial{t}^{2}} + \De\frac{\partial^{2}{p}}{\partial{t}\partial{r}} + \alpha^{2}\frac{\partial{u}}{\partial{t}}&=-\frac{\partial{p}}{\partial{r}}+\epsilon^{2}\frac{\partial^{2}{u}}{\partial{r}^{2}}+\epsilon^{2}\frac{1}{{r}}\frac{\partial{u}}{\partial{r}}+\frac{\partial^{2}{u}}{\partial{z}^{2}}-\epsilon^{2}\frac{{u}}{{r}^{2}},\label{eq:dimless_cont(a)}\\
\epsilon^{2}\De\,\alpha^{2}\frac{\partial^{2}{v}}{\partial{t}^{2}} + \De\frac{\partial^{2}{p}}{\partial{t}\partial{z}}+\epsilon^{2}\alpha^{2}\frac{\partial{v}}{\partial{t}}&=-\frac{\partial{p}}{\partial{z}}+\epsilon^{2}\left(\epsilon^{2}\frac{\partial^{2}{v}}{\partial{r}^{2}}+\epsilon^{2}\frac{1}{{r}}\frac{\partial{v}}{\partial{r}}+\frac{\partial^{2}{v}}{\partial{z}^{2}}\right),\label{eq:dimless_cont(b)}
\end{align}\end{subequations}
while the dimensionless form of the boundary conditions (\ref{eq:original BC_1palte0}) is
\begin{equation}
{u}(0)=0,\qquad{u}(1)=0,\qquad {v}(1)=\Re\left\{ \mathrm{i}\mathrm{e}^{\mathrm{i}{t}}\right\},\qquad {v}(0)=0.
\label{eq:original BC_1palte-1}
\end{equation}
Here, we define the Womersley number \citep{womersley_method_1955}, $\alpha^{2}=\hat{\rho}\hat{\omega} \hat{h}_{0}^{2}/\hat{\mu}$, the ratio of transient inertial forces to viscous forces, and the Deborah number, ${\De}=\hat{\lambda}_0\hat{\omega}$, which represents the ratio of the elastic relaxation time scale to the oscillation time scale.

Combining Eqs.~(\ref{eq:dimless_cont(a)}) and (\ref{eq:dimless_cont(b)})
at leading order in $\epsilon$, we obtain
\begin{equation}
\De\,\alpha^{2}\frac{\partial^{3}{u}}{\partial{t}^{2}\partial{z}}+\alpha^{2}\frac{\partial^{2}{u}}{\partial{t}\partial{z}}=\frac{\partial^{3}{u}}{\partial{z}^{3}}.
\label{eq:combined contr}
\end{equation}

\subsubsection{Solution}
\label{Section:Solution velocity field}
Assuming the following form of the solutions,  
\begin{subequations}
\label{eq:solution structure}
\begin{align}
{u}({r},{z},{t})&
=\Re\left\{ {\frac{r}{2}}{f}^{\prime}({z})\mathrm{e}^{\mathrm{i}{t}}\right\} ,\\
{v}({z},{t})&=-\Re\left\{ {f}({z})\mathrm{e}^{\mathrm{i}{t}}\right\} ,
\end{align}\label{eq:velocity form}\end{subequations}
and substituting them into Eqs.~(\ref{eq:combined contr}), we find that $f$ satisfies the following ordinary differential equation: 
\begin{equation}
(\mathrm{i}-\De)\alpha^{2}{f}^{\prime\prime}({z})-{f}^{(4)}({z})=0
\label{eq:ODE for f}
\end{equation}
subject to the boundary conditions from  Eq.~(\ref{eq:original BC_1palte-1}):
\begin{equation}
{f}^{\prime}(0)=0,\qquad {f}^{\prime}(1)=0,\qquad {f}(1)=-\mathrm{i},\qquad {f}(0)=0.
\label{eq:BC of 1wall}
\end{equation}

We note that the solution structure \eqref{eq:solution structure} imposes a particular behaviour of the fluid at $\hat{r}=\hat{R}$. In practice one should apply appropriate conditions, which specify, for example, how the free surface moves or conditions on a reservoir into which the fluid enters beyond $r=1$. However, these edge effects will play a small role in the overall behaviour as a result of the lubrication approximation that we have made along with the assumption of small wall oscillations, and thus \eqref{eq:solution structure} provides a suitable representation of the fluid flow.

The solution to Eq.~(\ref{eq:ODE for f}) subject to \eqref{eq:BC of 1wall} is 
\begin{equation}
{f}({z})=-\frac{\mathrm{ie}^{\mathrm{\Gamma}(1-{z})}\left(\mathrm{e}^{\mathrm{\Gamma}({z}-1)}\left(\Gamma{z}-\mathrm{e}^{\Gamma{z}}+\mathrm{e}^{\mathrm{\Gamma}}(\Gamma{z}-1)+1\right)+1\right)}{\mathrm{e}^{\mathrm{\Gamma}}(\Gamma-2)+\Gamma+2},
\end{equation} 
where $\Gamma=\alpha\sqrt{\mathrm{i}-\De}$. Thus, the dimensionless velocity components are
\begin{subequations}
\label{eq:u_1_v_1wall}
\begin{align}
{u}({r},{z},{t})&=-\Re\left\{ {\frac{r}{2}}\mathrm{e}^{\mathrm{i}{t}}\frac{\mathrm{i}\Gamma\left(\text{e}^{\Gamma}-\text{e}^{\Gamma{z}}-\text{e}^{\Gamma-\Gamma{z}}+1\right)}{\text{e}^{\Gamma}(\Gamma-2)+\Gamma+2}\right\} ,\label{eq:u_1wall}\\
{v}({z},{t})&=\Re\left\{ \frac{\mathrm{ie}^{\Gamma(1-{z})}\left(\mathrm{e}^{\Gamma({z}-1)}\left(\Gamma{z}-\mathrm{e}^{\mathrm{\Gamma}{z}}+\mathrm{e}^{\Gamma}(\Gamma{z}-1)+1\right)+1\right)}{\mathrm{e}^{\Gamma}(\Gamma-2)+\Gamma+2}\mathrm{e}^{\mathrm{i}{t}}\right\} .\label{eq:v_1wall}
\end{align}\end{subequations}

\subsection{Time-averaged scalar dispersion -- mass and heat transfer}
\label{sec:Diffusion}

\subsubsection{Governing equations\label{subsec:Equations}}

For the flow sketched in Fig.~\ref{fig:schematic} given by the solution in Eq.~\eqref{eq:u_1_v_1wall}, the advection--diffusion equation governing the mass and heat transfer (in the absence of mass/heat sources or sinks) can be expressed as:
\begin{equation}
\frac{\partial\hat{\Lambda}}{\partial \hat{t}} + \left(\frac{1}{\hat{r}}\frac{\partial(\hat{r}\hat{u}\hat{\Lambda})}{\partial \hat{r}}+\frac{\partial(\hat{v}\hat{\Lambda})}{\partial \hat{z}}\right)=\hat{D}\frac{1}{\hat{r}}\frac{\partial}{\partial \hat{r}}\left(\hat{r}\frac{\partial\hat{\Lambda}}{\partial \hat{r}}\right)+\hat{D}\frac{\partial^{2}\hat{\Lambda}}{\partial \hat{z}^{2}}.
\label{eq:heat tran. eq.2}
\end{equation}
where $\hat{D}$ is the mass or thermal diffusivity, and $\hat{\Lambda}$ represents either the concentration or the temperature for the cases of mass and heat transfer, respectively.  {Note that, for the equivalence between mass and heat transfer to hold, we assume the viscous dissipation in heat transfer is negligible. This assumption is valid when the Brinkman number, i.e., $ \sigma\hat{U}^{2}/(\hat{\Lambda}_{\mathrm{max}}-\hat{\Lambda}_{\mathrm{min}})\hat{c}_{p} = \mathrm{o}\left(\epsilon^{2}\right)$, where $\hat{\Lambda}_{\mathrm{max}}$,  $\hat{\Lambda}_{\mathrm{min}}$ and $\hat{c}_{p}$ represent the maximum temperature, minimum temperature and the specific heat capacity, respectively.} The walls of the plates are impermeable and thermally insulated, thus we have the following boundary conditions
\begin{equation}
\left.\frac{\partial\hat{\Lambda}}{\partial\hat{z}}\right|_{\hat{z}=0}=\left.\frac{\partial\hat{\Lambda}}{\partial\hat{z}}\right|_{\hat{z}=\hat{h}(\hat{t})}=0.
\label{eq:heat tran BC}
\end{equation}
To close the problem we also require boundary conditions at a fixed radial position, however, these are not necessary to determine the radially averaged governing equation; we consider a particular case in \textsection\ref{Section:Energy considerations}.    

By introducing the dimensionless variables from Eq.~\eqref{eq:nond_notation}, and the dimensionless concentration/temperature $\Lambda=(\hat{\Lambda}-\hat{\Lambda}_{\mathrm{min}})/(\hat{\Lambda}_{\mathrm{max}}-\hat{\Lambda}_{\mathrm{min}})$, Eqs.~\eqref{eq:heat tran. eq.2} and \eqref{eq:heat tran BC} become
\begin{subequations}\begin{align}
\alpha^{2}\sigma\frac{\partial\Lambda}{\partial{t}} + \epsilon \Pen\left(\frac{1}{{r}}\frac{\partial\left({r}{u}\Lambda\right)}{\partial{r}}+\frac{\partial\left({v}\Lambda\right)}{\partial{z}}\right)&=\frac{\epsilon^{2}}{{r}}\frac{\partial}{\partial{r}}\left({r}\frac{\partial\Lambda}{\partial{r}}\right)+\frac{\partial^{2}\Lambda}{\partial{z}^{2}},\label{eq:dispersion cont}\\[3mm]
\left.\frac{\partial\Lambda}{\partial{z}}\right|_{{z}=0}=\left.\frac{\partial\Lambda}{\partial{z}}\right|_{{z}=h}&=0.
\end{align}\label{eq:dispersion cont and bc}\end{subequations}
where $\Pen=\hat{h}_0 \hat{U}/\hat{D}$ is the P\'{e}clet number, and $\sigma=\hat{\mu}/\hat{\rho}\hat{D}$ is the Schmidt number in mass transfer or the  Prandtl number in heat transfer. 

\subsubsection{Multiple-time-scales analysis}

We now turn to the derivation of the time-averaged transport equation, for which we employ the technique of multiple time-scale homogenization \cite{kevorkian_multiple_1996}. There are three disparate time scales in this problem; first, the characteristic time for mass/heat diffusion in the transverse direction is:
\begin{equation}
\hat{t}_{0}=\mathcal{O} \left (\frac{\hat{h}_0^{2}}{\hat{D}}\right ).\label{eq:assumption of t0}
\end{equation}
We assume this time scale is comparable to the oscillation period, i.e., $\hat{t}_{0}=\mathcal{O}\left({1}/{\hat\omega}\right)$, which means the mass/heat diffusion in the transverse direction should equilibrate within several oscillation periods \citep{mei_homogenization_2010,ng_dispersion_2006}. 
The second time scale is the characteristic time  for advection in the streamwise direction:
\begin{equation}
 \hat{t}_{1}=\frac{\hat{t}_{0}}{\epsilon}=\mathcal{O} \left (\frac{\hat{R}}{\hat{U}}\right ).    
 \label{eq:assumption of t1}
\end{equation}
The third, and longest, time scale is for streamwise diffusion:
\begin{equation}
  \hat{t}_{2}=\frac{\hat{t}_{0}}{\epsilon^{2}}=\mathcal{O} \left (\frac{\hat{R}^{2}}{\hat{D}} \right).
  \label{eq:assumption of t2}
\end{equation} 

In the multiple-time-scales analysis, we assume that all variables are dependent on these three time scales \emph{independently} \citep{pagitsas_multiple_1986,kevorkian_multiple_1996}. Then, using the dimensionless versions of these time scales, the time derivative transforms as 
\begin{equation}
\frac{\partial}{\partial{t}}=\frac{\partial}{\partial{t}_{0}}+\frac{\text{d}{t}_{1}}{\text{d}{t}}\frac{\partial}{\partial{t}_{1}}+\frac{\text{d}{t}_{2}}{\text{d}{t}}\frac{\partial}{\partial{t}_{2}}=\frac{\partial}{\partial{t}_{0}}+\epsilon\frac{\partial}{\partial{t}_{1}}+\epsilon^{2}\frac{\partial}{\partial{t}_{2}}.
\label{eq:time_exp.}
\end{equation}

We follow \citet{fife_dispersion_1975} and expand $\Lambda$ as follows: 
\begin{equation}
\Lambda\sim\Lambda_{0}({r},{z},{t}_{1},{t}_{2})+\sum_{n=1}^{\infty}\epsilon^{n}\Lambda_{n}({r},{z},{t}_{0},{t}_{1},{t}_{2})+\sum_{n=0}^{\infty}\epsilon^{n}W_{n}({r},{z},{t}_{0}),
\end{equation}
where the $\Lambda_{n}$ are assumed to be fully developed terms that are harmonic functions of ${t}_{0}$, while the $W_{n}$ terms are assumed to be transient in ${t}_{0}$ and vanish as ${t}_{0}\to\infty$.

Next, the homogenization \citep{mei_applications_1996,mei_homogenization_2010} is performed (see details in Appendix~\ref{app:MMS}), which transforms Eq.~\eqref{eq:dispersion cont} into a homogenized  advection--diffusion--reaction equation for transport in the viscoelastic OSF:
\begin{equation}
\alpha^{2}\sigma\frac{\partial\Lambda_{0}}{\partial{t}_{2}}=\frac{1}{{r}}\frac{\partial}{\partial{r}}\left(\left(1+{D}_{{\rm eff}}\right){r}\frac{\partial\Lambda_{0}}{\partial{r}}-r{U}_{{\rm eff}}\Lambda_{0}\right)+{S}_{{\rm eff}}\Lambda_{0},
\label{eq:final_PDE_t2(2)}
\end{equation}
where 
\begin{subequations}
\label{eq:DUSeff}
\begin{align}
{D}_\mathrm{eff}(r)&=-\frac{{r}^{2}}{4}\Pen\, \Re\left\langle {f}'^{*}B_{\text{w}}\right\rangle ,
\label{eq:Deff}\\
{U}_\mathrm{eff}(r)&=-\frac{{r}}{2}\Pen\, \Re\left\{ \mathrm{i}B_{\text{w}}(1)\right\}  ,\label{eq:Ueff}\\
{S}_\mathrm{eff}&=-\Pen\, \Re\left\{ \mathrm{i}B_{\text{w}}(1)\right\}.
\label{eq:Seff}
\end{align}\end{subequations}
Here, the superscript $*$ represents the conjugate of a complex number, and we have further defined
\begin{align*}
B_{\text{w}}(z) & =\frac{\Gamma \Pen\,\mathrm{e}^{-{z}(\Gamma+2A_{1})}\left[A_{2}\left(-e^{A_{3}}+e^{\Gamma+A_{1}(2{z}+1)+\Gamma {z}}+e^{A_{1}(2{z}+1)+\Gamma {z}}-e^{{z}(\Gamma+2A_{1})}\right)+A_{4}\right]}{2\alpha^{2}\sigma A_{2}\left[e^{\Gamma}(\Gamma-2)+\Gamma+2\right]\left(e^{A_{1}}-1\right)},\\
A_{1} &= (-1)^{1/4}\alpha\sqrt{\sigma},\\
A_{2} &= \alpha^{2}\sigma+\mathrm{i}\Gamma^{2},\\
A_{3} &= \Gamma+2A_{1}{z}+\Gamma {z},\\
A_{4} &= \mathrm{i}A_{1}^{2}\left(e^{\Gamma+2A_{1}{z}+A_{1}}+e^{2A_{1}{z}+A_{1}+2\Gamma {z}}-e^{2{z}(\Gamma+A_{1})}-e^{\Gamma+2A_{1}{z}}\right)\nonumber\\
&\hspace{1cm}-\mathrm{i}A_{1}\Gamma\left(e^{({z}+1)(\Gamma+A_{1})}-e^{A_{1}{z}+A_{1}+\Gamma {z}}+e^{\Gamma+3A_{1}{z}+\Gamma {z}}-e^{\left(\Gamma+3A_{1}\right)z}\right).
\end{align*}

Using Eq.~(\ref{eq:time_exp.}), ${t}_2$ in Eq.~(\ref{eq:final_PDE_t2(2)}) can be replaced by the general time ${t}$, resulting in
\begin{equation}
\frac{\alpha^{2}\sigma}{\epsilon^{2}}\frac{\partial\Lambda_{0}}{\partial{t}}=\frac{1}{{r}}\frac{\partial}{\partial{r}}\left(\left(1+{D}_{{\rm eff}}\right){r}\frac{\partial\Lambda_{0}}{\partial{r}}-r{U}_{{\rm eff}}\Lambda_{0}\right)+{S}_{{\rm eff}}\Lambda_{0}.
\label{eq:final_PDE_generalt}
\end{equation}
Note that the above equation is valid on the long time scale, i.e., ${t}\gtrsim{t}_2$. {By recalling  $\hat{t}_0\ll\hat{t}_1\ll\hat{t}_2$ and substituting Eqs.~(\ref{eq:assumption of t0}) to (\ref{eq:assumption of t2}) into this inequality, we obtain a restriction on the range $\Pen$, i.e., $\epsilon\ll\Pen\ll {1}/{\epsilon}$, to ensure validity of our theory. Further, by substituting Eq.~(\ref{eq:assumption of t0}) into the assumption of $\hat{t}_{0}=\mathcal{O}\left({1}/{\hat\omega}\right)$, we obtain  $\alpha^2\sigma=\mathcal{O}\left({1}\right)$.}
 
Equation~(\ref{eq:final_PDE_generalt}) is the homogenized {effective} advection--diffusion--reaction equation for transport in the viscoelastic OSF, which embodies three effective mechanisms of mass/heat transfer comprising the dispersion process. 
The first one is the {effective} diffusion, with ${D}_{{\rm eff}}$ representing the {effective} streamwise diffusivity. 
As with molecular diffusion (Fick's law), the effective diffusive flux is also driven by the concentration gradient. 
The second is the {effective} advection, which indicates that mass/heat is carried along by an {effective} streamwise advection velocity ${U}_{{\rm eff}}$, representing a time-averaged drift velocity. 
When ${U}_{{\rm eff}}>0$, it is directed radially outwards; when ${U}_{{\rm eff}}<0$, it is directed radially inwards. 
The third mechanism of mass/heat transfer is an {effective} reaction term, releasing/absorbing solute or heat when ${S}_{{\rm eff}}>0$, or ${S}_{{\rm eff}}<0$, respectively. 
This reaction term arises as a result of the wall motion; we recall that no mass or heat is absorbed or emitted by the walls. 
The ratio $U_{\mathrm{eff}}/S_{\mathrm{eff}}={r}/{2}$, indicating that the effect of effective advection is much less significant than effective reaction when $r\ll1$, which is expected given the structure of the velocity field. 
When the effective advection and reaction work against the concentration/temperature gradient, and their effects on transport exceeds that of effective diffusion, the spreading of mass/heat will be inhibited. This is similar to the finding of a negative time-averaged dispersion coefficient \citep{chu2019dispersion}, with which the solute cloud will be compressed in the oscillatory flow. Additionally, these three effective mechanisms are not independent, as they all depend on the dimensions of the device, motion of the plate, and properties of the fluid. 

Comparing Eq.~(\ref{eq:final_PDE_generalt}) with the long-time equations obtained in similar treatments of transport in oscillatory pipe flow \citep{mei_homogenization_2010,ng_dispersion_2006,ramon_solute_2011,chu2019dispersion}, we find that the dispersion process in an OSF is different, in two aspects. First, the occurrence of the effective advection and reaction terms, beyond the usual dispersion term; these appear due to the first-order term at the boundary, which is embodied in $B_w(1)$. Second, ${D}_{{\rm eff}}$ varies in the streamwise (i.e., radial) direction, as well as ${U}_{{\rm eff}}$. The main reason is that the radial velocity component varies in the streamwise  direction, which is also a feature of steady radial flow with converging and diverging streamlines \citep{stone_dispersion_1999}.

\section{Results and discussion}
\subsection{Velocity field} \label{sec:dis_velocity}
\begin{figure}
\centering
 \begin{overpic}[scale=0.64,tics=5]{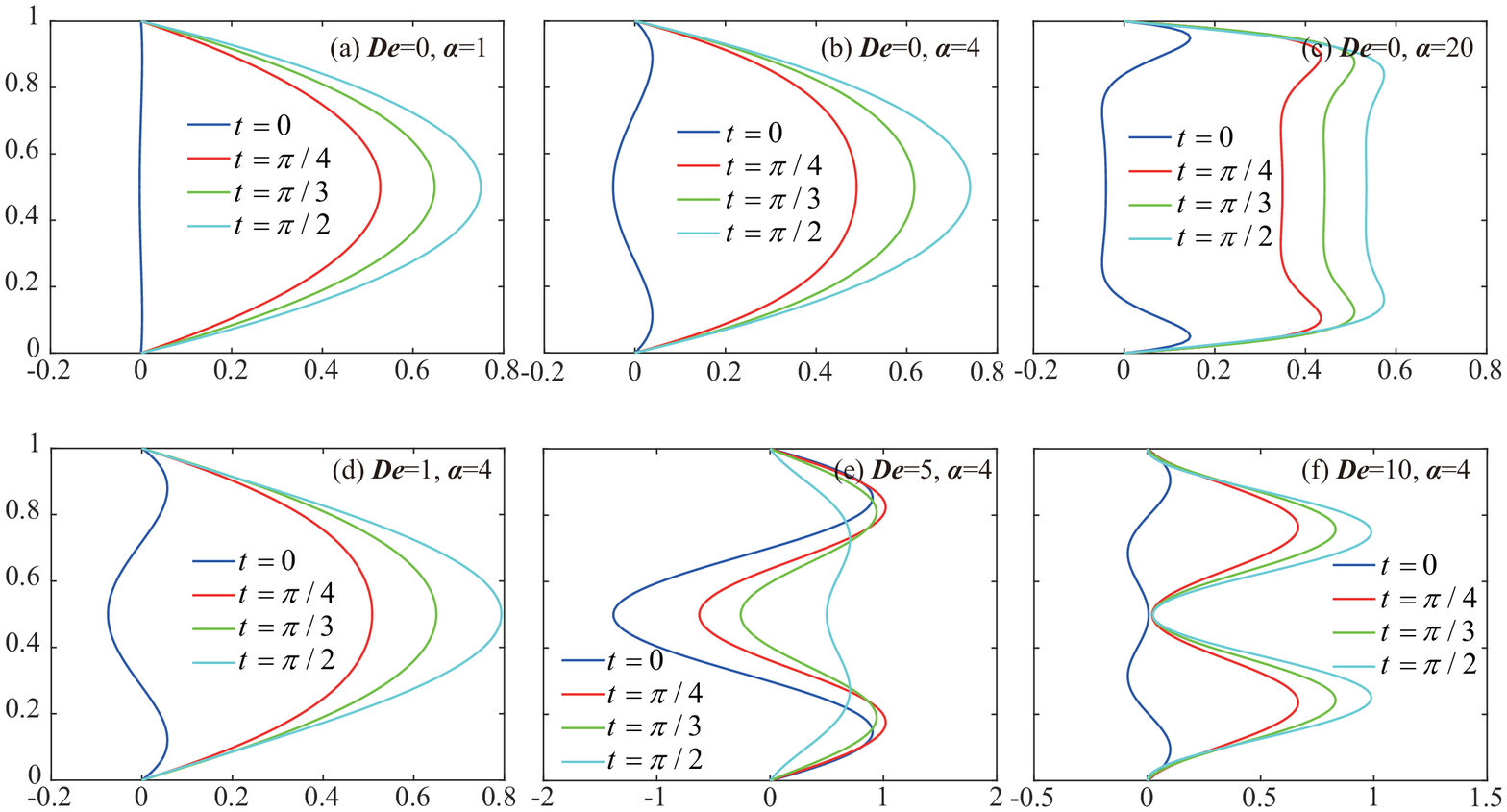}
\put(17.5,-1.5){$u$}
\put(50.55,-1.5){$u$}
\put(83,-1.5){$u$}
\put(17.5,27){$u$}
\put(51,27){$u$}
\put(83,27){$u$}
\put(-1.5,13){$z$}
\put(-1.5,41.5){$z$}
\end{overpic}
\caption{The dimensionless streamwise velocity ${u}$ variation across the gap height $z$ at ${r}=1$. Different values of $\De$ and $\alpha$ highlight the relative effects of viscoelasticity and oscillatory forcing.}
\label{fig: distr.of_u_1walls}
\end{figure}

In Fig.~\ref{fig: distr.of_u_1walls}, we present the profiles of the horizontal velocity ${u}$ at different times at the fixed cross-section ${r}=1$. Note that the value of $r$ is chosen arbitrarily for illustrative purposes; the shape of the velocity profile is unchanged by varying $r$. Comparing Figs.~\ref{fig: distr.of_u_1walls}(a) and (c), we observe that the Newtonian fluid's velocity profile ${u}$ becomes flatter near the centerline when $\alpha$ is relatively large. In this case, the forcing scale is smaller than the viscous time scale, so the effects of the oscillation of the top  plate are confined to near the wall during one oscillation period, and the fluid in the centerline is not influenced. 
The negligible pressure gradient in the ${z}$ direction for $De=0$ (see Eq.~(\ref{eq:dimless_cont(b)})) ensures the profile in near centerline remains flat. 
However, when $\alpha$ is small, the oscillations are slow enough so that viscosity affects the entire cross-section, as shown in (a). 
The velocity profiles shown in Figs.~\ref{fig: distr.of_u_1walls}(a) to (c) exhibit similar  variation with $\alpha$ as the oscillatory flow in a pipe \citep{womersley_elastic_1957,loudon_use_1998,ramon_solute_2011}.

Next, comparing Figs.~\ref{fig: distr.of_u_1walls}(d) and (f), we observe that the velocity has multiple local maxima, across the gap, when $\De\gg1$. The Deborah number $\De$ is the ratio of the elastic relaxation time to the oscillation time scale of the OSF, which is set by the plate motion. For a Newtonian fluid ($\De=0$), the flow generated by the movement of the plate is felt instantaneously throughout the entire gap. For a viscoelastic fluid ($\De>0$), on the other hand, the elasticity of the fluid acts as a restoring force that inhibits the transfer of momentum from the oscillatory wall. Thus, for larger values of $\De$, the influence of the motion of the plate lags behind the forcing, i.e., it takes ``longer'' to be felt throughout the domain. As a result, the fluid near the center lags behind the fluid near the wall, leading to the multiple peaks in the velocity distribution across the gap.

The nature of the fluid flow also changes dramatically at some specific values of $\De$ and $\alpha$. For instance, the phase of ${u}$ when $\De=5$ and $\alpha=4$ (Fig.~\ref{fig: distr.of_u_1walls}(e)) is very different from the other cases shown. This is attributed to the triggering of a `resonant' mode. To understand and quantify this effect, we introduce the spatio-temporal average of the longitudinal velocity
\begin{equation}
{u}_\mathrm{a}\left(r\right) =\frac{1}{2\pi}\int_{0}^{1}\int_{{0}}^{2\pi}\left|{u}\left(r,z,t\right)\right|\mathrm{d}{t}\mathrm{d}{z}.
\label{eq:uavg}
\end{equation}
The dependence of ${u}_\mathrm{a}$ on $\alpha$, for different values of $\De$ at radial position ${r}= 1$ is shown in Fig.~\ref{fig: u_vs_alpha,resonance}. It can be seen that, for a viscoelastic fluid ($\De>0$), there are peaks of ${u}_\mathrm{a}$ at specific values of $\alpha$. This phenomenon is caused by the elasticity of the fluid, hence we term it a \emph{viscoelastic resonance}. The corresponding value of $\alpha$ can be considered as the dimensionless resonant frequency. This effect has also been observed in the oscillatory pipe flow of a Maxwell fluid both experimentally \citep{castrejon-pita_experimental_2003} and numerically \citep{lambert_heat_2009}.  

We find that the variation of $u_\mathrm{a}$ is mainly determined by the denominator of $u_\mathrm{a}$, while its numerator is much less significant. We therefore introduce a parameter
\begin{equation}
\Theta(\alpha,\De)=\mathrm{e}^{\Gamma}\left(\Gamma-2\right)+\Gamma+2,\label{eq:denominator_of_u}
\end{equation}
representing the denominator of $u_\mathrm{a}$. As shown in Fig.~\ref{fig: u_vs_alpha,resonance}, the peaks of ${u}_\mathrm{a}$ coincide with the local minima of $\left|\Theta\right|$ when $\De=10$. 
We further compared the match between the peaks of  ${u}_\mathrm{a}$ and minima of $\left|\Theta\right|$ for $\De\leq100$, observing they match without exception for this range. 
Hence, it can be deduced that the resonance is governed by $\left|\Theta\right|$, which is a manifestation of the elasticity of the fluid. 
For a Maxwell fluid, as $\alpha$  increases, the amplitude of the oscillations of $\left|\Theta\right|$ decreases, while $\left|\Theta\right|$ continues to increase, leading to the decay of the oscillations. 
This effect can be interpreted as viscous dissipation overcoming the elastic restoring force. Consequently, the peaks of ${u}_\mathrm{a}$ gradually reduce with increasing $\alpha$ and eventually disappear beyond some critical value. 

\begin{figure}[t]
\centering\begin{overpic}[scale=1,tics=10]{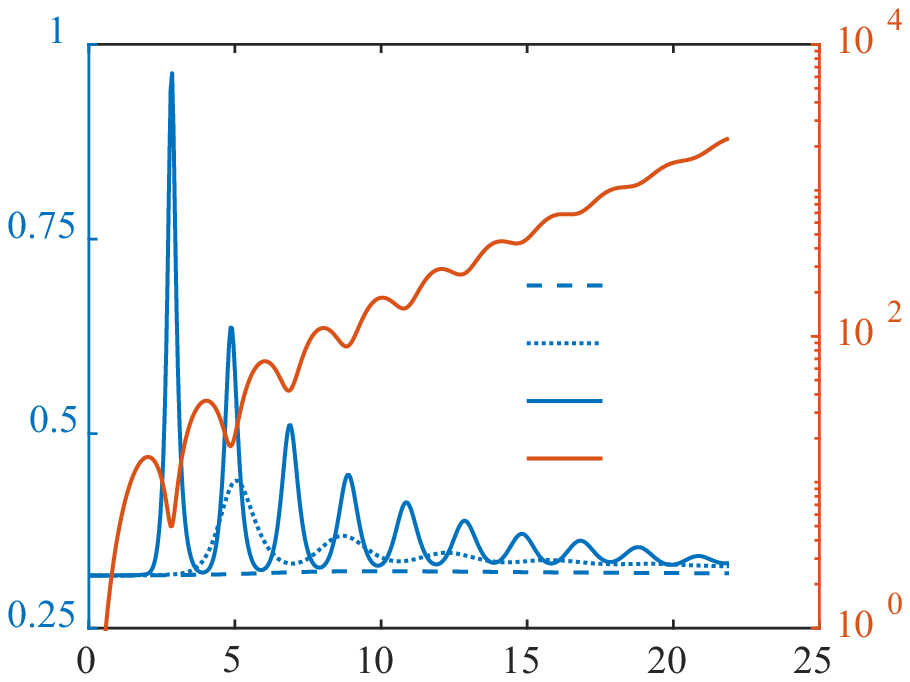}
\put(2,32){\large\color{bluegray}\begin{rotate}{90}${u}_\mathrm{a}$\end{rotate}}
\put(97,32){\color{cadmiumorange}\begin{rotate}{90}$\left|\Theta\right|$\end{rotate}}
\put(49,-2){$\alpha$}
\put(65,38.5){$\De=0$}
\put(65,33){$\De=3$}
\put(65,27){$\De=10$}
\put(65,21.5){$\De=10$}
\end{overpic}
\caption{The dependence of the amplitude of the $r$-direction velocity, ${u}_\mathrm{a}$ from Eq.~\eqref{eq:uavg}, and the absolute value of the denominator of ${u}$, $\left|\Theta\right|$ from Eq.~\eqref{eq:Predict_resonance}, on $\alpha$ at ${r}= 1$.}
\label{fig: u_vs_alpha,resonance}
\end{figure}

The resonant values of $\alpha$ satisfy $\partial |\Theta|/\partial \alpha = 0$, corresponding with local minima of $\left|\Theta\right|$. Specifically, beginning with Eq.~\eqref{eq:denominator_of_u}, and aided by numerical experimentation, we find the following approximation for the $\alpha$ value corresponding to the $n^{\textrm{th}}$ peak, \begin{equation}\label{eq:Predict_resonance}
\alpha^\ast_n \approx \frac{\pi \left(2n+1\right)}{\Im\left\{ \sqrt{\mathrm{i}-\De}\right\}},
\end{equation}
where $\Im\{\,\cdot\,\}$ denotes the imaginary part of a complex number, and $n \in \mathbb{N}^+$, such that the $\alpha$ corresponding to the first peak of ${u}_\mathrm{a}$ can be approximated with $n=1$, the second peak with $n=2$, and so on. 
Comparing the estimated values of $\alpha^\ast_n$ obtained from Eq.~(\ref{eq:Predict_resonance}), and the corresponding exact values in the range of $\De\leq100$ and $n\leq10$, we find that the deviations are quite small and gradually decrease as $\De$ increases. For instance, the error in $\alpha^\ast_1$ is 0.32 $(6.3\%)$, 0.21 $(5.4\%)$ and 0.04 $(4.7\%)$ for $\De=3$, $\De=5$ and $\De=100$, respectively. 

Equation~(\ref{eq:Predict_resonance}) also indicates that, as $\De$ is increased, the value of $\alpha^\ast_n$ decreases for fixed $n$, as does the interval between neighboring resonant peaks, as seen in Fig.~\ref{fig: u_vs_alpha,resonance}. Note that there is no visible peak for a Newtonian fluid when the value of $\alpha$ reaches $\alpha^\ast_1$, which is 13.33 according to Eq.~(\ref{eq:Predict_resonance}), in Fig.~\ref{fig: u_vs_alpha,resonance}, because the amplitude of fluctuation of $\left|\Theta\right|$ is too small to be compared with its absolute value. As $\De$  increases, the height of the peaks, especially the first peak, increases, presumably because for larger $\De$, the value of $\alpha^\ast_1$ decreases (see Eq.~(\ref{eq:Predict_resonance})), thus $\left|\Theta\right|$ becomes very small, leading to a very high peak of ${u}_\mathrm{a}$. This viscoelastic resonance is significant for the mass and heat transfer in the OSF, which is discussed in the following section.

\subsection{Time-averaged transport characteristics} \label{sec:dis_diffusion}
 
\subsubsection{Effective dispersion, advection and reaction}
In this section, we first discuss how ${D}_\mathrm{eff}$, ${U}_\mathrm{eff}$ and ${S}_\mathrm{eff}$, at the cross-section of ${r}=1$, vary with the Womersley $\alpha$, Deborah $\De$ and Schmidt (or Prandtl) $\sigma$ numbers. As with the discussion of velocity field in Sec.~\ref{sec:dis_velocity}, the value of $r$ is chosen for illustrative purposes, and this choice does not affect the general characteristics identified. To facilitate the discussion, we introduce the tidal displacement, which represents the cross-sectionally averaged longitudinal distance traversed by a fluid particle under the OSF velocity field over half a period. This quantity is calculated through the spatio-temporal average of the velocity magnitude:
\begin{equation}
   \Delta r\left(r\right)=\frac{\delta}{2\epsilon}\int_{0}^{1}\int_{0}^{2\pi}\left|u\left(r,z,t\right)\right|\mathrm{d}t\mathrm{d}z=\frac{\pi\delta}{\epsilon}u_{\textrm{a}}\left(r\right),
\label{eq:Deltar}
\end{equation}
where $u_\textrm{a}$ is defined by Eq.~\eqref{eq:uavg} and must be evaluated numerically.

In addition, following \citet{kurzweg_tuning_1986}, who scaled the effective diffusivity (the dispersion coefficient) for an oscillatory pipe flow by the product of the frequency and the square of the tidal displacement, we define the re-scaled diffusivity, advection velocity and reaction coefficients for the OSF:
\begin{equation}
    \lambda_{\mathrm{D}}=\frac{{D}_\mathrm{eff}}{\Delta{r}^{2}},\qquad     \lambda_{\mathrm{U}}=\frac{{U}_\mathrm{eff}}{\Delta{r}^{2}},\qquad     \lambda_{\mathrm{S}}=\frac{{S}_\mathrm{eff}}{\Delta{r}^{2}}.
\label{eq:lambda_D,U,S}
\end{equation}
After this re-scaling, $\lambda_{\mathrm{D}}$, $\lambda_{\mathrm{U}}$ and $\lambda_{\mathrm{S}}$ are independent of the dimensionless displacement of the upper plate, i.e., $\delta$. Since $\Delta r$ depends on $\delta$, $\lambda_{\mathrm{D}}$, $\lambda_{\mathrm{U}}$ and $\lambda_{\mathrm{S}}$ can also be independent of $\Delta r$ if $u_\mathrm{a}$ and $\epsilon$ stays fixed  according to Eq.~(\ref{eq:Deltar}). 

\begin{figure}[t]
\centering\begin{overpic}[scale=0.74,tics=5]{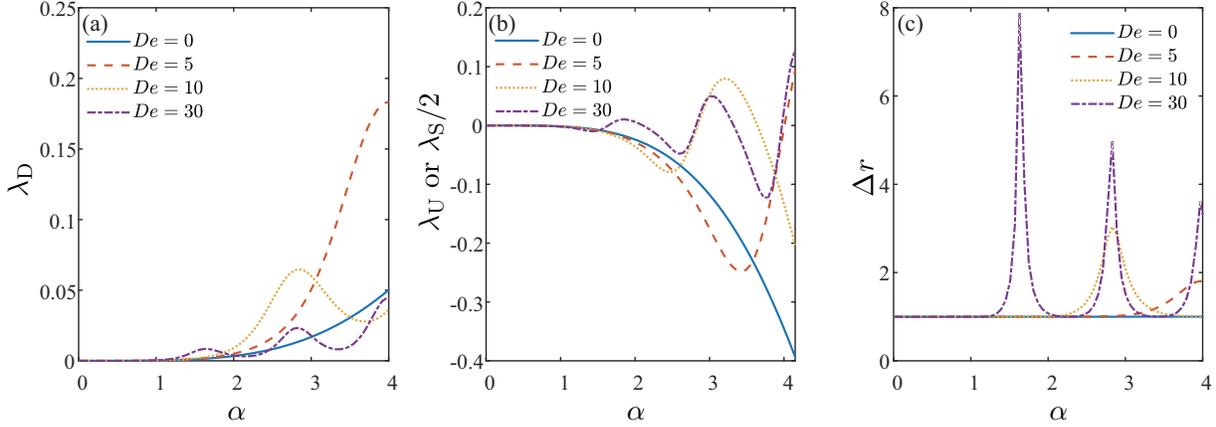}
\put(16,-2){$\alpha$}
\put(50,-2){$\alpha$}
\put(86,-2){$\alpha$}
\put(-1,17){\begin{rotate}{90}$\lambda_\mathrm{D}$\end{rotate}}
\put(34,14){\begin{rotate}{90}$\lambda_\mathrm{U}$ or ${\lambda_\mathrm{S}}/{2}$\end{rotate}}
\put(71,17){\begin{rotate}{90}$\Delta r$\end{rotate}}
\end{overpic}
\caption{The dependence of $\lambda_\mathrm{D}$, $\lambda_\mathrm{U}$ (or ${\lambda_\mathrm{S}}/{2}$) and $\Delta{r}$ on $\alpha$ for different values of $\De$, at ${r}=1$ with $\sigma=0.6$. 
$\Delta{r}$ is obtained with $\delta=\epsilon$. The value of $\Pen$ is determined by $\Pen=\alpha^2\sigma \delta / \epsilon$. Note that $\lambda_\mathrm{U}={\lambda_\mathrm{S}}/{2}$ for $r=1$.}
\label{fig:Deff_vs_De_1wall}
\end{figure}

It is important to note that $\lambda_\mathrm{D}$, $\lambda_\mathrm{U}$ and $\lambda_\mathrm{S}$ strongly depend on the values of $\alpha$ and $\De$, as shown in Fig.~\ref{fig:Deff_vs_De_1wall}. Specifically, optimal values of $\alpha$ have local maxima in terms of $\lambda_\mathrm{D}$, which depend on the value of $\De$. 
For a given range of $\alpha$, increasing $\De$ will generate more peaks, but not necessarily a higher value of $\lambda_\mathrm{D}$. 
As an example, $\lambda_\mathrm{D}$ has a maximum at $\alpha= 4$, when $\De=30$, but this maximum value is not higher than the one for $\De=0$. For a Maxwell fluid ($De>0$), the maximum values of  $\lambda_\mathrm{D}$ increase with $\alpha$, in accordance with the oscillatory pipe flow in which the effective diffusivity is proportional to the frequency \citep{kurzweg_tuning_1986}. 

The peaks in Fig.~\ref{fig:Deff_vs_De_1wall}(a) are the resonant values of $\lambda_\mathrm{D}$, because the corresponding values of $\alpha$ appear to closely match the resonant frequencies, i.e., the values of $\alpha$ corresponds to the peaks in Fig.~\ref{fig:Deff_vs_De_1wall}(c). 
In other words, the resonance sharply increases both the values of $\lambda_\mathrm{D}$ and $\Delta{r}$. Hence, the locations of peaks in Fig.~\ref{fig:Deff_vs_De_1wall}(a) and (c) can be approximated by Eq.~(\ref{eq:Predict_resonance}), although with slight deviations. From Figs.~\ref{fig:Deff_vs_De_1wall}(a) and (c), we observe opposite trends of the height of peaks in $\lambda_\mathrm{D}$ and $\Delta{r}$ with increasing $\alpha$. Accordingly,
the value of $D_\mathrm{eff}$ might be dominated by $\Delta{r}$ for a Maxwell fluid with relatively large $\De$ at small $\alpha$ or, conversely, by $\lambda_\mathrm{D}$ for a weakly viscoselastic fluid (small $\De$) but at large $\alpha$. This interplay determines whether the OSF leads to enhanced diffusion (dispersion) or diminished transport.
For instance, when $\alpha=1.64$ with $\De=30$, the value of ${D}_\mathrm{eff}$ is two orders of magnitude larger than that with the absence of resonance, which is mainly attributed to $\Delta{r}$, which is increased by a factor of eight by resonance. 

The fact that the ratio $\lambda_{\mathrm{U}}/\lambda_{\mathrm{S}}=r/2$ indicates that transport due to the effective advection is much less significant than that due to effective reaction when $r\ll1$. In Fig.~\ref{fig:Deff_vs_De_1wall}(b), for a Newtonian fluid ($De=0$), $\lambda_{\mathrm{U}}$ and $\lambda_{\mathrm{S}}$ are negative, indicating that the effective advection is always directed from the edge of the plate towards the center, and the effective reaction absorbs mass/heat. For a Maxwell fluid, $\lambda_{\mathrm{U}}$ and $\lambda_{\mathrm{S}}$ fluctuate around 0, with the amplitude increasing with $\alpha$, indicating that the effect of the effective advection and reaction, i.e., carrying mass/heat from high-concentration region to low-concentration region or along the opposite direction, is very sensitive to the oscillatory forcing of the system (quantified by $\alpha$). 

\begin{figure}[t]
\centering\begin{overpic}[scale=0.8,tics=5]{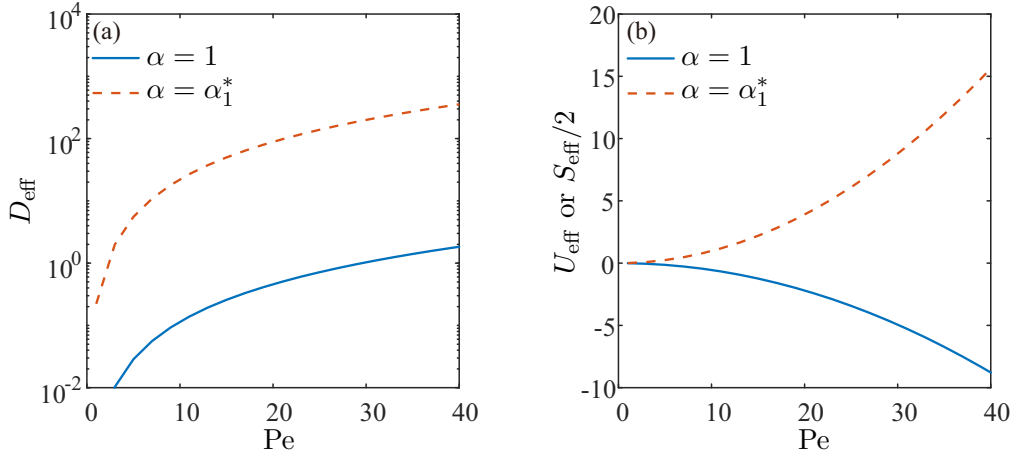}
\put(11,37.5){$\alpha=1$}
\put(11,34.1){$\alpha=\alpha^\ast_1$}
\put(66,37.5){$\alpha=1$}
\put(66,34.1){$\alpha=\alpha^\ast_1$}
\put(23,-2){$\Pen$}
\put(78,-2){$\Pen$}
\put(-1,21){\begin{rotate}{90}$D_\mathrm{eff}$\end{rotate}}
\put(55,16){\begin{rotate}{90}$U_\mathrm{eff}$ or ${{S}_\mathrm{eff}}/{2}$\end{rotate}}
\end{overpic}
\caption{The dependence of ${D}_\mathrm{eff}$ and ${U}_\mathrm{eff}$ (or ${{S}_\mathrm{eff}}/{2}$) on $\Pen$,
at ${r}=1$ when $\De=30$ and $\sigma=3$. Two different values of $\alpha$ are shown: $\alpha=1$ (no resonance) and $\alpha^\ast_1=1.64$ (the first resonant mode). The values of $\alpha^\ast_n$ were obtained numerically. Note that ${U}_\mathrm{eff}={{S}_\mathrm{eff}}/{2}$  for $r=1$.}
\label{fig:eta_vs_Pe}
\end{figure}

Figure~\ref{fig:eta_vs_Pe} shows the dependence of ${D}_\mathrm{eff}$, ${U}_\mathrm{eff}$ and ${S}_\mathrm{eff}$ on $\Pen$, at the first resonant mode, second resonant mode and with the absence of resonance. 
Importantly, ${D}_\mathrm{eff}$ is significantly increased by the existence of a viscoelastic resonance, because both $\Delta{r}$ and $\lambda_\mathrm{D}$ reach the peak values according to Fig.~\ref{fig:Deff_vs_De_1wall}. 
The strongest enhancement exists at the first resonant mode ($\alpha^\ast_1=1.64$), where the value of ${D}_\mathrm{eff}$ is around 200 times larger the corresponding value (at the same P\'{e}clet number $\Pen$) but in the absence of a resonance. 
Similarly, ${U}_\mathrm{eff}$ and ${S}_\mathrm{eff}$ also increase, although less significantly. Moreover, we can achieve larger ${D}_\mathrm{eff}$, ${U}_\mathrm{eff}$ and ${S}_\mathrm{eff}$ by increasing the displacement amplitude of the upper plate, which is embodied in $\Pen$. 

\begin{figure}[t]
\centering\begin{overpic}[scale=0.74,tics=3]{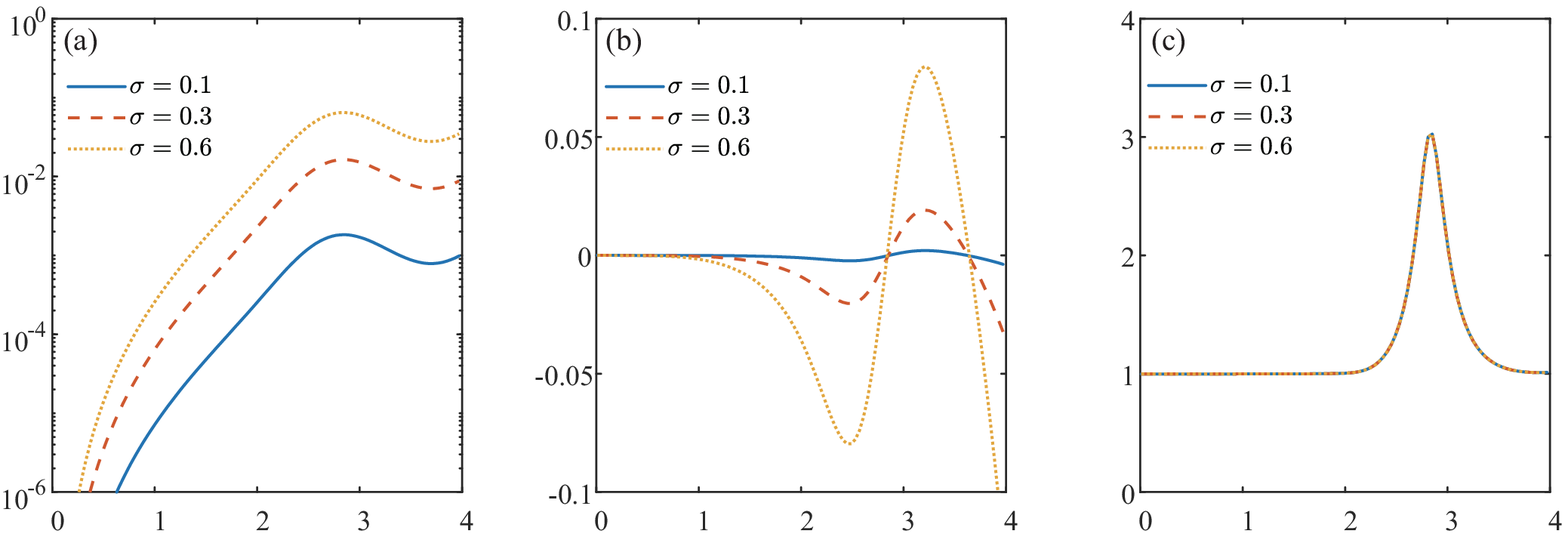}
\put(16,-2){$\alpha$}
\put(50,-2){$\alpha$}
\put(86,-2){$\alpha$}
\put(-1,17){\begin{rotate}{90}$\lambda_\mathrm{D}$\end{rotate}}
\put(34,14){\begin{rotate}{90}$\lambda_\mathrm{U}\ \mathrm{or}\ {\lambda_\mathrm{S}}/{2}$\end{rotate}}
\put(71,17){\begin{rotate}{90}$\Delta r$\end{rotate}}
\end{overpic}
\caption{The dependence of $\lambda_\mathrm{D}$, $\lambda_\mathrm{U}$ (or ${\lambda_\mathrm{S}}/{2}$) and $\Delta{r}$ on $\alpha$ with different values of $\sigma$, at ${r}=1$ when $\De=10$. The value of $\Delta{r}$ is obtained by choosing a constant value of displacement of the upper plate, which satisfies $\delta=\epsilon$. The value of $\Pen$ is determined by $\Pen=\alpha^2\sigma \delta / \epsilon$. Note that $\lambda_\mathrm{U}={\lambda_\mathrm{S}}/{2}$ for $r=1$.  }
\label{fig:Deff_vs_Sc_1wall-1}
\end{figure}

It is important to note, however, large values of ${D}_\mathrm{eff}$, ${U}_\mathrm{eff}$ and ${S}_\mathrm{eff}$ do not necessarily lead to the enhancement of mass/heat transfer: the  transport in the OSF is controlled jointly by the interaction of these three effective (homogenized) transport mechanisms. 
The effective diffusion (dispersion) always enhances transport, but whether the effective advection or reaction enhance or impede transport depends on whether they are directed along or against the concentration gradient, which is determined by the sign of $\lambda_{\mathrm{U}}$ (or ${U}_\mathrm{eff}$) and $\lambda_{\mathrm{S}}$ (or ${S}_\mathrm{eff}$) and the boundary conditions. 
If the effective advection and reaction work against the concentration gradient (so that the corresponding mass flux exceeds that given by effective diffusion), then the transport is diminished. This indicates that, to enhance the mass/heat transfer, the value of $\alpha$ and $\De$ should be carefully chosen, so that the effective advection and reaction assist in carrying the mass/heat from regions of high temperature or concentration to regions of low temperature or concentration. 

From Fig.~\ref{fig:Deff_vs_Sc_1wall-1}, we see that $\lambda_\mathrm{D}$, $\lambda_\mathrm{U}$ and $\lambda_\mathrm{S}$ grow with $\sigma$. However, this does not necessarily lead to the enhancement of effective mass/heat transfer because the effective advection and reaction might inhibit the transport if mass/heat is carried against the temperature or concentration gradient. Moreover, the increase of $\sigma=\hat{\mu}/\hat{\rho}\hat{D}$ requires a decrease of the molecular diffusivity $\hat{D}$ (for fixed $\hat{\mu}/\hat{\rho}$) or an increase of the kinematic viscosity $\hat{\nu}=\hat{\mu}/\hat{\rho}$ (for fixed $\hat{D}$), with the former generally reducing mixing, while the latter increases the energy needed to drive the OSF. Note that $\Delta r$ is a kinematic quantity independent of $\sigma$. Therefore, all the curves in Fig.~\ref{fig:Deff_vs_Sc_1wall-1}(c) overlap.

\subsubsection{Energy considerations for transport enhancement}
\label{Section:Energy considerations}
We now turn to consider the energy consumption associated with the OSF proposed herein for enhancing heat and/or mass transfer. To this end, in this section, we discuss the mass/heat flux, consumed power to drive the upper plate, and the efficiency in a specific example. 

\begin{figure}[t]
\centering\includegraphics[width=0.6\textwidth]{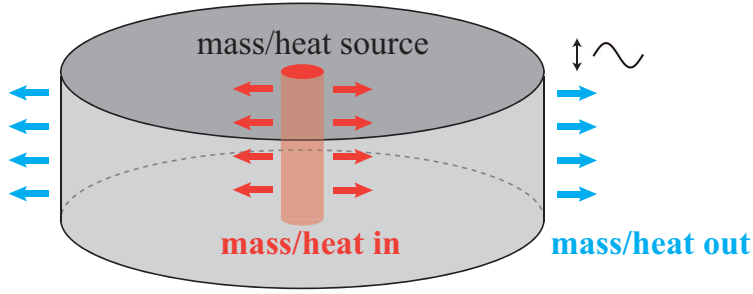}\caption{An example of an OSF configuration for enhancing mass or heat transfer.}
\label{fig: osf_example}
\end{figure}

Consider that a mass/heat source with a dimensionless radius of 0.1 is immersed in the center of an OSF device, as shown in Fig~\ref{fig: osf_example}. We assume it has no influence on the local velocity field. This source ensures $\Lambda_{0}=1$ at ${r}=0.1$. Suppose also that there is a mass/heat sink at $r=1$, which we assume not to affect the fluid flow but which forces $\Lambda_{0}=0$ at ${r}=1$. 
The top and bottom plates are impermeable and insulated walls, thus the amount of mass/heat injected from the source must equal absorbed by the sink at ${r}=1$. With the boundary condition at $r=1$, Eq.~(\ref{eq:final_PDE_generalt}) is reduced into a diffusion equation, thus the mass/heat flux transported from the source to the sink can be calculated by
\begin{equation}
\dot{m}^\ast=-\left(1+D_{{\rm eff}}\right)\left.\frac{\partial\Lambda_{0}}{\partial r}\right|_{r=1}. \label{eq:m_dot}
\end{equation}

The power required to drive the upper plate is found to be
\begin{equation}
\Tilde{W}_\mathrm{T}=\frac{1}{2\pi}\int_{0}^{2\pi}\left.v\right|_{z=1}\tilde{F}_{T}\,\mathrm{d}t=-\Re\left\{ \frac{\ri\pi f^{\prime\prime\prime}(1)}{16\left(1+\ri\,\De\right)}\right\} ,\label{eq:W_T}
\end{equation}
where $\tilde{F}_{T}$ is the excess normal force on the top plate (see Appendix~\ref{app:energy} for details).  We can then define a metric for the energy consumption per unit mass/heat flux, or efficiency,
\begin{equation}
\eta=\frac{\dot{m}^\ast}{\Tilde{W}_\mathrm{T}}.
\end{equation}

\begin{figure}[t]
\centering\begin{overpic}[scale=0.8,tics=5]{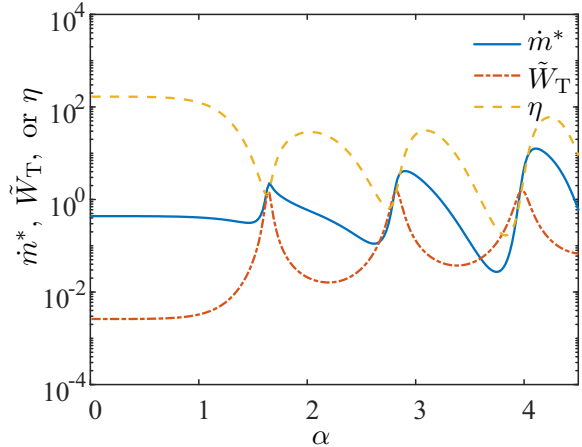}
\put(90,58){$\eta$}
\put(90,70){$\dot{m}^\ast$}
\put(90,63){$\Tilde{W}_\mathrm{T}$}
\put(50,-3){$\alpha$}
\put(-1,30){\begin{rotate}{90}$\dot{m}^\ast,\ \Tilde{W}_\mathrm{T},\ \mathrm{or}\ \eta$\end{rotate}}
\end{overpic}
\caption{The dependence of $\dot{m}^\ast$, $\eta$ and $\Tilde{W}_\mathrm{T}$ on $\alpha$ at steady state for a Maxwell fluid with $\De=30$ and $\sigma=0.6$. The amplitude of the oscillation of the upper plate is $\delta=10\epsilon$. The value of $\Pen$ is determined by $\Pen=\alpha^2\sigma \delta / \epsilon$. }
\label{fig: mass_flux}
\end{figure}

The dependence of $\dot{m}^\ast$, $\Tilde{W}_\mathrm{T}$ and $\eta$ on $\alpha$, for a Maxwell fluid with $\De=30$, is shown in Fig.~\ref{fig: mass_flux}. 
For $\alpha<1$, $\dot{m}^\ast$ is approximately constant (around 0.44) because $D_\mathrm{eff} \ll 1$ (see Fig.~\ref{fig:Deff_vs_De_1wall}), indicating that the transport is mainly driven by molecular diffusion (or conduction), which is independent of $\alpha$. 
For $\alpha>1$, $\dot{m}^\ast$ oscillates about the value corresponding to molecular diffusion (or pure conduction). 
According to Eq.~(\ref{eq:m_dot}), $\dot{m}^\ast$ is
influenced by both $D_\mathrm{eff}$ and ${\partial\Lambda_{0}}/{\partial r}$ at $r=1$. The value of $D_\mathrm{eff}$  is always positive and reaches a near-peak value at the resonant mode (see Fig.~\ref{fig:Deff_vs_De_1wall}), indicating that the effective diffusion always enhances the mass or heat transfer, especially at resonance. However, the dependence of  ${\partial\Lambda_{0}}/{\partial r}$ on $\alpha$ is more complex, as this strongly depends on $U_\mathrm{eff}$, $S_\mathrm{eff}$, and the boundary conditions.

In this example, when the values of $U_\mathrm{eff}$ and $S_\mathrm{eff}$ are positive, the value of ${\partial\Lambda_{0}}/{\partial r}$ is increased and thus the mass flux is enhanced; otherwise, the mass flux is diminished. 
For instance, the value of $\dot{m}^\ast$ is 30 times larger than that under pure molecular diffusion (which happens when $\alpha\rightarrow0$) when $\alpha\approx$4.1,  where both $U_\mathrm{eff}$ and $S_\mathrm{eff}$ are positive. 
On the other hand, the value of $\dot{m}^\ast$ is only $6\%$ of that under pure molecular diffusion when $\alpha\approx3.7$, because both  $U_\mathrm{eff}$ and $S_\mathrm{eff}$ are negative and near the local minimum values. 
The value of $\Tilde{W}_\mathrm{T}$ rises as $\alpha$ is increased, and reaches the local maximum exactly at the resonant mode, because the abrupt increase of velocity results in a significantly higher viscous loss. 
The efficiency $\eta$ is large when $\alpha<1$ because molecular diffusion (or pure conduction) does not consume energy; $\eta$ drops when $\alpha$ approaches the first resonant frequency because of the increased $\Tilde{W}_\mathrm{T}$. In this example, we see that the mass/heat transfer in such OSF configurations strongly depends on the Womersley number $\alpha$, which quantifies the oscillatory forcing of the flow. Thus, by choosing $\alpha$ suitably, we can either significantly increase the mass/heat flux and improve transport, or diminish the flux of mass/heat transfer to a value even lower than that of the pure molecular diffusion or pure conduction.

\section{Conclusion}
\label{sec:Conclusion}

In this work, we have investigated Taylor--Aris dispersion related to mass and/or heat transfer in an oscillatory squeeze flow of a viscoelastic (Maxwell) fluid. First, we derived the expression for the post-transient velocity for both Newtonian and Maxwell fluids. Then, we used the method of homogenization (a multiple-time-scale analysis) to determine the effective mass and/or heat transport equation on the long time scale. Specifically, we showed that transport is governed by an effective one-dimensional advection--diffusion--reaction equation. In doing so, we identified the three effective transport mechanisms: the effective diffusive spreading mass/heat along the concentration gradient, the effective advection carrying mass/heat along or against the concentration gradient, and the effective reaction releasing or absorbing solute.

Importantly, we found resonances in the viscoleastic oscillatory squeeze flow, which lead to an abrupt rise in the velocity, when the Womersley number (a dimensionless measure of the forcing frequency and unsteady inertia of the fluid) reaches specific values. 
We showed that these resonant values can be estimated by a simple expression involving the Deborah number. 
For a linearly viscoelastic Maxwell fluid, we found that there are multiple values of the Womersley number that may trigger the resonance, but the effect of resonance diminishes gradually with the Womersley number. 
We also observed that the mass/heat transfer in this viscoelastic fluid flow can be either enhanced or inhibited, depending on the value of the Womersley number. 
On the one hand, when the dimensionless frequency, i.e., the Womersley number, reaches the resonant frequency, the effective diffusion can be sharply enhanced, which accelerates the spread of mass/heat in the fluid. For instance, for a viscoelastic Maxwell fluid, we found that the effective diffusivity at the first resonant frequency can be 200 times larger than the corresponding one in the absence of a resonance. 
On the other hand, we found that the values of the effective advection and reaction coefficients oscillate about zero with increasing amplitudes as the Womersley number increases. 
Notably, these two effective mechanisms may enhance or inhibit mass/heat transfer in this oscillatory viscoelastic flow, depending on their signs, which is determined by the Womersley number and the fluid properties (including the Deborah number and the Schmidt or Prandtl numbers). Therefore, to enhance the mass transfer, the value of the Womersley number should be carefully chosen, so that the effective advection and reaction help to carry solute from regions of high temperature or solute concentration to regions of low concentration.

The results presented in this work suggest a new approach to enhance the mass/heat transfer using oscillatory squeeze flow. Further investigation could focus on the transport problem for large oscillation amplitudes, going beyond the assumptions inherent to the symptotic nature of the homogenization theory employed here.
\section*{Acknowledgements}
This research was partially supported by grant 2018/17 from the Israel
Science Foundation (ISF). R.Y.\ was supported, in part, by a fellowship from the Israel Council for Higher Education. I.C.C.\ acknowledges the donors of the American Chemical Society Petroleum Research Fund for partial support under ACS PRF award \# 57371-DNI9. I.M.G.\ gratefully acknowledges support from the Royal Society through a University Research Fellowship.
I.C.C., I.M.G., and G.Z.R.\ acknowledge the Collaborative Workshop Initiative (CWI) for providing a platform to instigate this research, as well as H.A.\ Stone for many insightful discussions on Taylor dispersion.

\bibliography{Dispersion4}


\appendix
\section*{Appendix}

\section{Derivation of the effective advection--diffusion equation using the homogenization method}
\label{app:MMS}

At $\mathcal{O}(1)$, Eqs.~(\ref{eq:dispersion cont and bc}) can be simplified
as
\begin{subequations}
\label{eq:appenddix1}\begin{align}
\alpha^{2}\sigma\frac{\partial\Lambda_{0}}{\partial{t}_{0}}&=\frac{\partial^{2}\Lambda_{0}}{\partial{z}^{2}},\\[2mm]
\left.\frac{\partial\Lambda_{0}}{\partial{z}}\right|_{{z}=0}&=\left.\frac{\partial\Lambda_{0}}{\partial{z}}\right|_{{z}=h}=0.
\end{align}\end{subequations}
If we seek the developed periodic solution, then $\Lambda_{0}$ is independent of the short time scale ${t}_{0}$. From the system \eqref{eq:appenddix1} we find $\Lambda_{0}$ is independent of ${z}$. Hence, $\Lambda_{0}=\Lambda_{0}({t}_{1},{t}_{2},{r})$.

At $\mathcal{O}(\epsilon)$, Eq.~(\ref{eq:dispersion cont and bc}) gives 
\begin{subequations}\begin{align}
\alpha^{2}\sigma\left(\frac{\partial\Lambda_{0}}{\partial{t}_{1}}+\frac{\partial\Lambda_{1}}{\partial{t}_{0}}\right)+\Pen\left[\frac{1}{{r}}\frac{\partial}{\partial{r}}({r}{u}\Lambda_{0})+\frac{\partial}{\partial{z}}({v}\Lambda_{0})\right] &= \frac{\partial^{2}\Lambda_{1}}{\partial{z}^{2}},\label{eq:dispersion_order_epsilon-1}\\[2mm]
\left.\frac{\partial\Lambda_{1}}{\partial{z}}\right|_{{z}=0}=\left.\frac{\partial\Lambda_{1}}{\partial{z}}\right|_{{z}=h} &= 0.
\label{eq:dispersion_order_epsilon-1_B}
\end{align}\end{subequations}

Next, we introduce the temporal and spatial averaging operators \begin{equation}
\overline{(\cdot)}=\frac{1}{2\pi}\int_{{t}}^{{t}+2\pi}(\cdot)\,\mathrm{d}{t}_{0},\qquad \langle\cdot\rangle=\frac{1}{h}\int_{0}^{h}(\cdot)\,\mathrm{d}{z} \simeq \int_{0}^{1}(\cdot)\,\mathrm{d}{z}.
\end{equation}
Note that $h\simeq1$ since $\hat{a}/\hat{h}_{0}\ll1$. 
Time averaging  Eqs.~(\ref{eq:dispersion_order_epsilon-1}) and (\ref{eq:dispersion_order_epsilon-1_B}) gives 
\begin{subequations}\begin{align}
\alpha^{2}\sigma\frac{\partial\Lambda_{0}}{\partial{t}_{1}} &= \frac{\partial^{2}\overline{\Lambda_{1}}}{\partial{z}^{2}}\label{eq:order_epsilon_2},\\
\left.\frac{\partial\overline{\Lambda_{1}}}{\partial{z}}\right|_{{z}=0}=\left.\frac{\partial\overline{\Lambda_{1}}}{\partial{z}}\right|_{{z}=h} &= 0\label{eq:order_epsilon_2_BC},
\end{align}\end{subequations}
where we have used the fact that  $\overline{{u}}=\overline{{v}}=0$ because $u$ and $v$ are time-harmonic functions (see Eq.~\eqref{eq:velocity form}) and $\overline{\Lambda_{0}}=\Lambda_{0}$ and $\langle\Lambda_{0}\rangle=\Lambda_{0}$ because $\Lambda_{0}$ is independent of ${t}_{0}$ and ${z}$, while $\overline{\partial\Lambda_{1}/\partial{t}_{0}}=0$.

Applying spatial average to Eq.~(\ref{eq:order_epsilon_2}) (and from the related boundary conditions shown in Eq.~(\ref{eq:order_epsilon_2_BC})), we have
\begin{equation}
\frac{\partial\Lambda_{0}}{\partial{t}_{1}}=0.
\label{eq:dC0dt1-1}
\end{equation}
With Eq.~(\ref{eq:dC0dt1-1}), Eq.~(\ref{eq:dispersion_order_epsilon-1})
can be simplified to
\begin{align}
\alpha^{2}\sigma\frac{\partial\Lambda_{1}}{\partial{t}_{0}}+\Pen\left[\frac{1}{{r}}\frac{\partial}{\partial{r}}({r}{u}\Lambda_{0})+\frac{\partial}{\partial{z}}({v}\Lambda_{0})\right] &= \frac{\partial^{2}\Lambda_{1}}{\partial{z}^{2}}.
\label{eq:dispersion_order_epsilon2-1}
\end{align}
According to the simplification $\Lambda_{0}=\Lambda_{0}\left({t}_{1},{t}_{2},{r}\right)$ and Eq.~(\ref{eq:velocity form}), we have 
\begin{align}
\frac{\partial}{\partial{z}}\left({v}\Lambda_{0}\right)=-\Lambda_{0}\Re\left\{ {f}^{\prime}({z})\mathrm{e}^{\mathrm{i}{t}}\right\}, \label{eq:dispersion_order_epsilon_dvCdz-1}\\
\frac{1}{{r}}\frac{\partial}{\partial{r}}({r}{u}\Lambda_{0})=\Lambda_{0}\Re\left\{ {f}^{\prime}({z})\mathrm{e}^{\mathrm{i}{t}}\right\} +\frac{r}{2}\Re\left\{ {f}^{\prime}({z})\mathrm{e}^{\mathrm{i}{t}}\right\} \frac{\partial\Lambda_{0}}{\partial{r}}.
\label{eq:order_epsilon_1/rd(ruC0)-1}
\end{align}
Substituting Eqs.~(\ref{eq:dispersion_order_epsilon_dvCdz-1}) and (\ref{eq:order_epsilon_1/rd(ruC0)-1})
into Eq.~(\ref{eq:dispersion_order_epsilon2-1}), we have 
\begin{equation}
\alpha^{2}\sigma\frac{\partial\Lambda_{1}}{\partial{t}_{0}}+\Pen\,\Re\left\{ {f}^{\prime}({z})\mathrm{e}^{\mathrm{i}{t}}\right\} \frac{{r}}{2}\frac{\partial\Lambda_{0}}{\partial{r}}=\frac{\partial^{2}\Lambda_{1}}{\partial{z}^{2}},
\label{eq:ODE of C1-1}
\end{equation}
which indicates that the solution of $\Lambda_{1}$ is of the form 
\begin{equation}
\Lambda_{1}={r}\frac{\partial\Lambda_{0}}{\partial{r}}\Re\left\{ B_{\mathrm{w}}({z})e^{\mathrm{i}{t}}\right\}.
\label{eq:form of C1-1}
\end{equation}
Substituting Eq.~(\ref{eq:form of C1-1}) into Eq.~(\ref{eq:ODE of C1-1}), we have 
\begin{align}
\frac{\rd^{2}B_{w}({z})}{\rd{z}^{2}}-\frac{1}{2}\Pen{f}^{\prime}({z})-\mathrm{i}\alpha^{2}\sigma B_{w}({z})&=0,\\
\left.\frac{\rd B_{w}}{\rd z}\right|_{{z}=0}=\left.\frac{\rd  B_{w}}{\rd z}\right|_{{z}=h}&=0.\label{eq:ODE for Bw-1}
\end{align}

At $\mathcal{O}(\epsilon^{2})$, Eqs.~(\ref{eq:dispersion cont and bc}) give 
\begin{subequations}\begin{align}
\alpha^{2}\mathrm{\sigma}\left(\frac{\partial\Lambda_{2}}{\partial{t}_{0}}+\frac{\partial\Lambda_{1}}{\partial{t}_{1}}+\frac{\partial\Lambda_{0}}{\partial{t}_{2}}\right)+\Pen\left[\frac{1}{{r}}\frac{\partial}{\partial{r}}({r}{u}\Lambda_{1})+\frac{\partial}{\partial{z}}\left({v}\Lambda_{1}\right)\right]&=\frac{1}{{r}}\frac{\partial}{\partial{r}}\left({r}\frac{\partial\Lambda_{0}}{\partial{r}}\right)+\frac{\partial^{2}\Lambda_{2}}{\partial{z}^{2}},
\label{eq:dispersion_order_epsilon^2-1}\\
\left.\frac{\partial\Lambda_{2}}{\partial{z}}\right|_{{z}=0}=\left.\frac{\partial\Lambda_{2}}{\partial{z}}\right|_{{z}=h}&=0.
\end{align}\end{subequations}
Equations~(\ref{eq:form of C1-1}) and (\ref{eq:dC0dt1-1}) give 
\begin{equation}
\frac{\partial\Lambda_{1}}{\partial{t}_{1}}=0.
\end{equation}
Taking the cross-sectional average of Eq.~(\ref{eq:dispersion_order_epsilon^2-1}), we have 
\begin{multline}
\alpha^{2}\sigma\left(\frac{\partial\left\langle \Lambda_{2}\right\rangle }{\partial{t}_{0}}+\frac{\partial\Lambda_{0}}{\partial{t}_{2}}\right)+\Pen\left[\left\langle \frac{1}{{r}}\frac{\partial}{\partial{r}}\left({r}{u}\Lambda_{1}\right)\right\rangle + \left({v}\Lambda_{1}\right)\Big|_{{z}=0}^{{z}=1}\right]
=\frac{1}{{r}}\frac{\partial}{\partial{r}}\left({r}\frac{\partial\Lambda_{0}}{\partial{r}}\right)+{\left\langle \frac{\partial^{2}\Lambda_{2}}{\partial{z}^{2}}\right\rangle }.
\label{eq:dispersion_order_epsilon^22-1}
\end{multline}

It can be shown that
\begin{align}
\left\langle \frac{1}{{r}}\frac{\partial}{\partial{r}}\left({r}{u}\Lambda_{1}\right)\right\rangle =\frac{\partial}{\partial{r}}\left\langle {u}\Lambda_{1}\right\rangle +\frac{1}{{r}}\left\langle {u}\Lambda_{1}\right\rangle, \\
\left\langle \frac{\partial^{2}\Lambda_{2}}{\partial{z}^{2}}\right\rangle =\left.\frac{\partial\Lambda_{2}}{\partial{z}}\right|_{{z}=0}^{{z}=1}=0,
\end{align}
then Eq.~(\ref{eq:dispersion_order_epsilon^22-1}) can be written as
\begin{equation}
\alpha^{2}\sigma\left(\frac{\partial\left\langle \Lambda_{2}\right\rangle }{\partial{t}_{0}}+\frac{\partial\Lambda_{0}}{\partial{t}_{2}}\right)+\Pen\left[\frac{\partial}{\partial{r}}\left\langle {u}\Lambda_{1}\right\rangle +\frac{1}{{r}}\left\langle {u}\Lambda_{1}\right\rangle +\left({v}\Lambda_{1}\right)\Big|_{{z}=0}^{{z}=1}\right]=\frac{1}{{r}}\frac{\partial}{\partial{r}}\left({r}\frac{\partial\Lambda_{0}}{\partial{r}}\right).
\label{eq:dispersion_order_epsilon^22-2}
\end{equation}
Taking the time average of Eq.~(\ref{eq:dispersion_order_epsilon^22-2}), we have
\begin{equation}
\alpha^{2}\sigma\frac{\partial\Lambda_{0}}{\partial{t}_{2}}+\Pen\left[\frac{\partial}{\partial{r}}\left\langle \overline{{u}\Lambda_{1}}\right\rangle +\frac{1}{{r}}\left\langle \overline{{u}\Lambda_{1}}\right\rangle +\overline{{v}\Lambda_{1}}\Big|_{{z}=0}^{{z}=1}\right]=\frac{1}{{r}}\frac{\partial}{\partial{r}}\left({r}\frac{\partial\Lambda_{0}}{\partial{r}}\right)\label{eq:epsilon^2_stavg-1}.
\end{equation}
Additionally, from Eqs.~(\ref{eq:velocity form}) and (\ref{eq:form of C1-1}), we reach
\begin{align}
\left\langle \overline{{u}\Lambda_{1}}\right\rangle &=\frac{{r}^{2}}{4}\frac{\partial\Lambda_{0}}{\partial{r}}\Re\left\langle {f}'({z})^{*}B_{{\rm w}}({z})\right\rangle, \label{eq:epsilon^2_vC1_avg-1}\\
\left.\overline{{v}\Lambda_{1}}\right|_{{z}=0}^{{z}=1}&=-\frac{{r}}{2}\frac{\partial\Lambda_{0}}{\partial{r}}\Re\left\{ {\rm i}B_{{\rm w}}\left(1\right)\right\}. \label{epsilon^2_uC1_avg-1}
\end{align}
Substituting Eqs.~(\ref{eq:epsilon^2_vC1_avg-1}) and (\ref{epsilon^2_uC1_avg-1}) into Eq.~(\ref{eq:epsilon^2_stavg-1}), we have
\begin{multline}
\alpha^{2}\sigma\frac{\partial\Lambda_{0}}{\partial{t}_{2}}=\frac{\partial^{2}\Lambda_{0}}{\partial{r}^{2}}\left[1-\frac{{r}^{2}}{4}\Re\left\langle {f}'({z})^{*}B_{{\rm w}}({z})\right\rangle \Pen\right] + 
\frac{\partial\Lambda_{0}}{\partial{r}}\left[
\frac{1}{{r}}\left(1-\frac{{r}^{2}}{4}\Re\left\langle {f}'({z})^{*}B_{{\rm w}}({z})\right\rangle \Pen\right) \right.\\ \left. -\frac{{r}}{2}\Re\left\langle {f}'({z})^{*}B_{{\rm w}}({z})\right\rangle \Pen+\frac{{r}}{2}\Re\left\{ {\rm i}B_{{\rm w}}(1)\right\} \Pen\right].
\label{eq:dispersion_final-1}
\end{multline}

Introducing the dimensionless effective diffusivity, velocity and reaction terms shown in Eq.~(\ref{eq:DUSeff}),
Eq.~(\ref{eq:dispersion_final-1}) can be arranged into Eq.~(\ref{eq:final_PDE_t2(2)}).

\section{Derivation of the component-wise momentum equations}
\label{app:DDt}

The material derivative, assuming axisymmetry, is 
\begin{equation}
\frac{\mathrm{D}}{\mathrm{D}\hat{t}}=\frac{\partial}{\partial \hat{t}}+\hat{u}\frac{\partial}{\partial \hat{r}}+\hat{v}\frac{\partial}{\partial \hat{z}}.
\end{equation}
Using the dimensionless variables from  Eq.~(\ref{eq:nond_notation}), we have
\begin{equation}
\ensuremath{\frac{\mathrm{D}}{\mathrm{D}t}=\frac{\partial}{\partial t}+\delta u\frac{\partial}{\partial r}+\delta v\frac{\partial}{\partial z}}.
\end{equation}
We recall that $\delta={\hat{a}}/{\hat{h}_{0}}\ll1$ is the dimensionless displacement amplitude of the upper plate. Neglecting the small terms of $\mathcal{O}(\delta)$, we obtain
\begin{equation}
\frac{\mathrm{D}}{\mathrm{D}\hat{t}} \simeq \frac{\partial}{\partial \hat{t}}\,.
\label{eq:materialD}
\end{equation}
Next, combining Eq.~(\ref{eq:maxwell}) and Eq.~(\ref{eq:momentum_0}) and acknowledging Eq.~(\ref{eq:materialD}), we have 
\begin{equation}
\hat{\lambda}_0\hat{\rho}\frac{\partial^{2}\hat{\bm{u}}}{\partial \hat{t}^{2}}+\hat{\lambda}_0\frac{\partial}{\partial \hat{t}}\left(\nabla \hat{p}\right)+\hat{\rho}\frac{\partial\hat{\bm{u}}}{\partial \hat{t}}=-\nabla \hat{p}+\hat{\mu}\nabla^{2}\hat{\bm{u}}.
\label{eq:momentum_dim}
\end{equation}
From Eq.~(\ref{eq:momentum_dim}), we obtain the component-wise momentum equations given in Eqs.~(\ref{eq:momentum1_dim}) and (\ref{eq:momentum2_dim}).

\section{Energy and power requirement}
\label{app:energy}

In the fully periodic regime, the dimensionless pressure is of the form
\begin{equation}
{p}=\Re\left\{ {p}_{0}({r},{z})\mathrm{e}^{\mathrm{i}{t}}\right\}. \label{eq:dim_p}
\end{equation}
Substituting Eqs.~(\ref{eq:velocity form}) and (\ref{eq:dim_p}) into Eqs.~(\ref{eq:dimless_cont(a)})  at
leading order in $\epsilon$, we have
\begin{equation}
\left(1+\ri\,\De\right)\frac{\partial{p}_{0}}{\partial{r}}=\left(\De-\ri\right)\alpha^{2}{\frac{r}{2}}{f}^{\prime}({z})+{\frac{r}{2}}{f}^{\prime\prime\prime}({z}),\label{eq:3}
\end{equation}
Integrating Eq.~(\ref{eq:3}), we obtain
\begin{equation}
{p}_{0}({r},{z})=\frac{{r}^{2}}{4\left(1+\ri\,\De\right)}\left[\left(\De-\ri\right)\alpha^{2}{f}^{\prime}({z})+{f}^{\prime\prime\prime}({z})\right]+G(z),\label{eq:p_0_0}
\end{equation}
where $G\left(z\right)$ is an arbitrary function of integration. We define an \emph{excess} pressure as
\begin{equation}
\tilde{p}=\Re\left\{ \tilde{p}_{0}({r},{z})\mathrm{e}^{\mathrm{i}{t}}\right\}, \label{eq:dim_p4}
\end{equation}
where the amplitude $\tilde{p}_0$ is relative to the background pressure amplitude $\left.p_0\right|_{r=1}$, i.e.,
\begin{equation}
\tilde{p}_0=p_0-\left.p_0\right|_{r=1}.\label{eq:excess_p}
\end{equation}
Substituting Eq.~(\ref{eq:p_0_0}) into Eq.~(\ref{eq:excess_p}), we obtain
\begin{equation}
\tilde{p}_{0}({r},{z})=\frac{{r}^{2}}{4\left(1+\ri\,\De\right)}\left[\left(\De-\ri\right)\alpha^{2}{f}^{\prime}({z})+{f}^{\prime\prime\prime}({z})\right]+G_2(z)\label{eq:p_0},
\end{equation}
where $G_2=G-\left.p_0\right|_{r=1}$. 
Solving for $G_2$ with the boundary condition of $\left.\tilde{p}_0\right|_{r=1}=0$, we obtain
\begin{equation}
\tilde{p}_0(r,z)=\frac{r^{2}-1}{4\left(1+\ri\,\De\right)}\left[\left(\De-\ri\right)\alpha^{2}f^{\prime}(z)+f^{\prime\prime\prime}(z)\right].
\label{eq:ex_p}
\end{equation}

From Eq.~(\ref{eq:maxwell}), the component-wise
equation for the stress tensor $\hat{\tau}_{zz}$ can be expressed as
\begin{equation}
\hat{\lambda}_{0}\left(\frac{\mathrm{\partial}\hat{\tau}_{zz}}{\mathrm{\partial}\hat{t}}\right)=\hat{\mu}\left(\nabla\hat{v}+\nabla\hat{v}^{T}\right)-\hat{\tau}_{zz}=2\hat{\mu}\frac{\partial\hat{v}}{\partial\hat{z}}-\hat{\tau}_{zz}.
\end{equation}
With the non-dimensionalization in Eq.~(\ref{eq:nond_notation}) and $\hat{\tau}_{zz}={\tau}_{zz}{\hat{\mu} \hat{a}\hat{\omega}}/{\epsilon^{3}\hat{R}}$, we have
\begin{equation}
\frac{\De}{\epsilon^{2}}\frac{\mathrm{\partial}{\tau}_{zz}}{\mathrm{\partial}{t}}=2\frac{\partial{v}}{\partial{z}}-\frac{1}{\epsilon^{2}}{\tau}_{zz}.
\end{equation}
By writing ${\tau}_{zz}=\Re\left\{ {\tau}_{0,zz}(z)\mathrm{e}^{\mathrm{i}{t}}\right\} $, we obtain
\begin{equation}
{\tau}_{zz}=-\Re\left\{ \frac{2\epsilon^{2}}{1+\ri\,\De}{f}^{\prime}({z})\mathrm{e}^{\mathrm{i}{t}}\right\}. \label{eq:tau}
\end{equation}

The scaled normal force on the top plate is found to be (see \cite{phan-thien_viscoelastic_1983}):
\begin{equation}
{F}_\mathrm{T}=2\pi\int_{0}^{1}{\sigma}_{zz}\,{r}\,\mathrm{d}{r}=2\pi\int_{0}^{1}\left.\left({\tau}_{zz}-{p}\right)\right|_{{z}=1}{r}\,\mathrm{d}{r}.
\label{eq:dim_F}
\end{equation}
Recalling that ${f}^{\prime}(1)=0$ from Eq.~(\ref{eq:BC of 1wall}), Eq.~(\ref{eq:dim_F}) can be simplified as 
\begin{equation}
{F}_\mathrm{T}=-2\pi\int_{0}^{1}\left.{p}\right|_{{z}=1}{r}\,\mathrm{d}{r}.
\label{eq:dim_F2}
\end{equation}
We define an excess normal force
\begin{equation}
\Tilde{F}_\mathrm{T}=-2\pi\int_{0}^{1}\left.\Tilde{p}\right|_{{z}=1}{r}\,\mathrm{d}{r}.
\label{eq:dim_F3}
\end{equation}
Taking $\Tilde{F}_\mathrm{T}=\Re\left\{ \Tilde{F}_\mathrm{0,T}\mathrm{e}^{\mathrm{i}{t}}\right\}$ and substituting Eq.~(\ref{eq:dim_p4})  into Eq.~(\ref{eq:dim_F3}), we obtain
\begin{equation}
\Tilde{F}_\mathrm{0,T}=\frac{\pi f^{\prime\prime\prime}(1)}{8\left(1+\ri\,\De\right)}.
\label{eq:F_0}
\end{equation}

Hence, the power needed to drive the upper plate is 
\begin{equation}
\tilde{W}=\frac{1}{2\pi}\int_{0}^{2\pi}\left.v\right|_{z=1}\tilde{F}_{T}\,\mathrm{d}t=-\Re\left\{ \frac{\ri\pi f^{\prime\prime\prime}(1)}{16\left(1+\ri\,\De\right)}\right\}.
\end{equation}

\end{document}